\documentclass[aps, prx,twocolumn,longbibliography,superscriptaddress,amsmath,amssymb,floatfix]{revtex4-2}

\usepackage{amssymb}
\usepackage{graphicx}
\usepackage{amsmath}
\usepackage{color}
\usepackage{mathrsfs}
\usepackage{float}
\usepackage{indentfirst}
\usepackage{mathrsfs}
\usepackage{float}
\usepackage{indentfirst}
\usepackage{textcomp}
\usepackage{comment}
\usepackage{mathtools}
\usepackage{natbib,hyperref}
\usepackage{soul}  
\usepackage{bm}

\definecolor{deepskyblue}{rgb}{0.0, 0.75, 1.0}
\begin{document}
\title{From the Shastry-Sutherland model to the $J_1$-$J_2$ Heisenberg model}

\author{Xiangjian Qian}
\affiliation{Key Laboratory of Artificial Structures and Quantum Control (Ministry of Education),  School of Physics and Astronomy, Shanghai Jiao Tong University, Shanghai 200240, China}

\author{Rongyi Lv}
\affiliation{Key Laboratory of Artificial Structures and Quantum Control (Ministry of Education),  School of Physics and Astronomy, Shanghai Jiao Tong University, Shanghai 200240, China}

\author{Jong Yeon Lee} \thanks{jongyeon@illinois.edu}
\affiliation{Department of Physics and Institute of Condensed Matter Theory, University of Illinois at Urbana-Champaign, Urbana, Illinois 61801, USA}

\author{Mingpu Qin} \thanks{qinmingpu@sjtu.edu.cn}
\affiliation{Key Laboratory of Artificial Structures and Quantum Control (Ministry of Education),  School of Physics and Astronomy, Shanghai Jiao Tong University, Shanghai 200240, China}

\affiliation{Hefei National Laboratory, Hefei 230088, China}

\date{\today}


\begin{abstract}
    We propose a generalized Shastry-Sutherland model which bridges the Shastry-Sutherland model and the $J_1$-$J_2$ Heisenberg model. By employing large scale Density Matrix Renormalization Group and Fully Augmented Matrix Product State calculations, combined with careful finite-size scaling, we find the phase transition between the plaquette valence bond state (PVBS) and Neel anti-ferromagnetic (AFM) phase in the pure Shastry-Sutherland model is a weak first one. This result indicates the existence of an exotic tri-critical point in the PVBS to AFM transition line in the phase diagram, as the transition in the $J_1$-$J_2$ Heisenberg model was previously determined to be continuous. We determine the location of the tri-critical point in the phase diagram at which first-order transition turns to continuous. Our generalized Shastry-Sutherland model provides not only a valuable platform to explore exotic phases and phase transitions but also more realistic description of Shastry-Sutherland materials like SrCu$_2$(BO$_3$)$_2$. 
\end{abstract}

\maketitle
{\em Introduction--}
The exploration of quantum phases and phase transitions in strongly correlated systems remains a central theme in condensed matter physics, offering profound insights into the fundamental behaviors of matter and the exotic phenomena these systems exhibit~\cite{doi:10.1126/science.adc9487,Balents2010,Savary_2017,doi:10.1126/science.1091806,senthil2023deconfined,PhysRevLett.49.957,PhysRevLett.52.1583,PhysRevLett.53.722,PhysRevLett.128.146803,PhysRevLett.126.015301,Ludwig_2011,PhysRevLett.126.160604,RevModPhys.78.17,PhysRevLett.127.097002,xu2023coexistence}. Among the various models that have captivated the attention of physicists, the Shastry-Sutherland (SS) model \cite{SRIRAMSHASTRY19811069} and the $J_1$-$J_2$ Heisenberg model stand out as paradigms for studying quantum magnetism and the intricate interplay of competing interactions.

\begin{figure*}[t]
    \includegraphics[width=70mm]{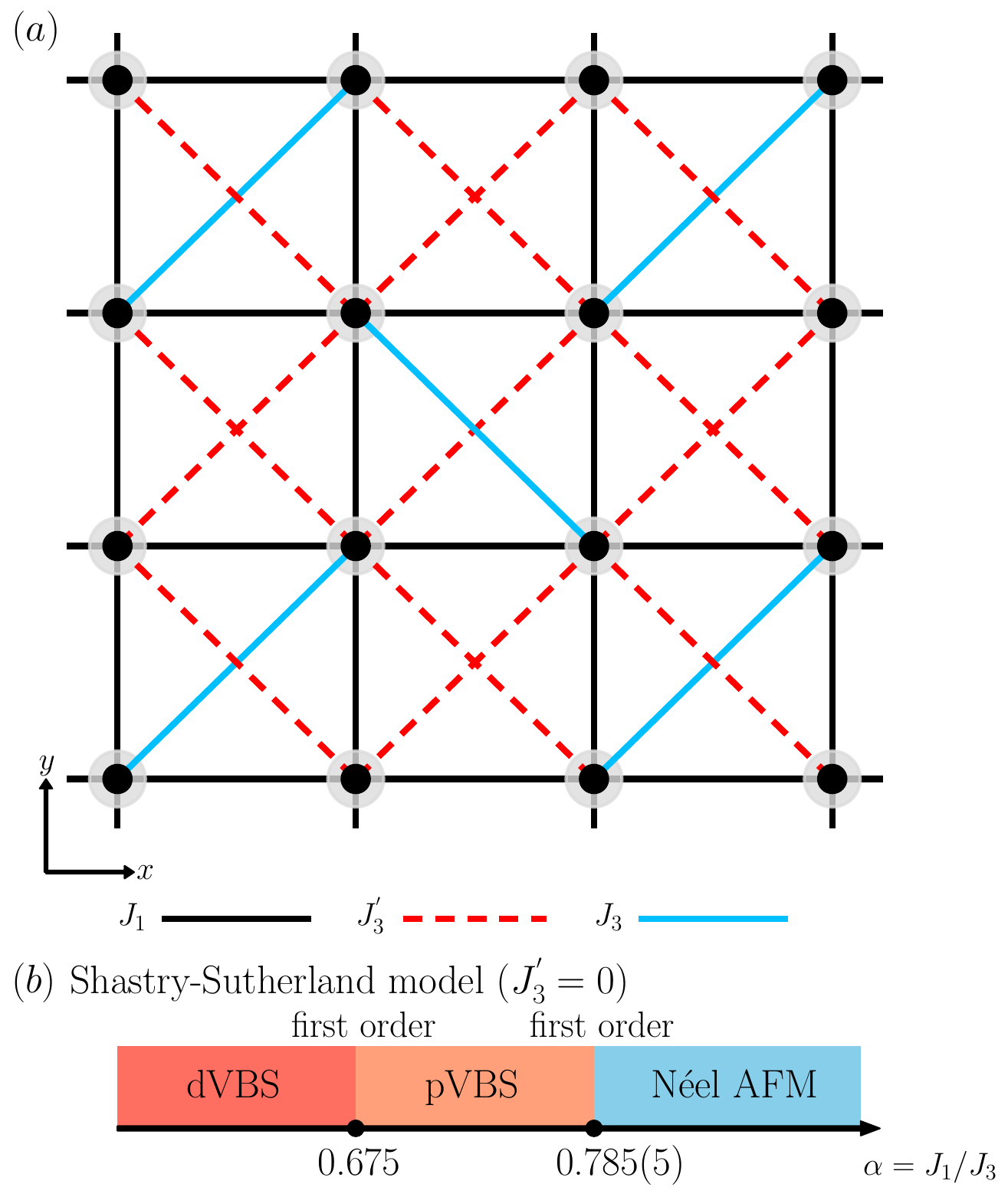}
    \includegraphics[width=82mm]{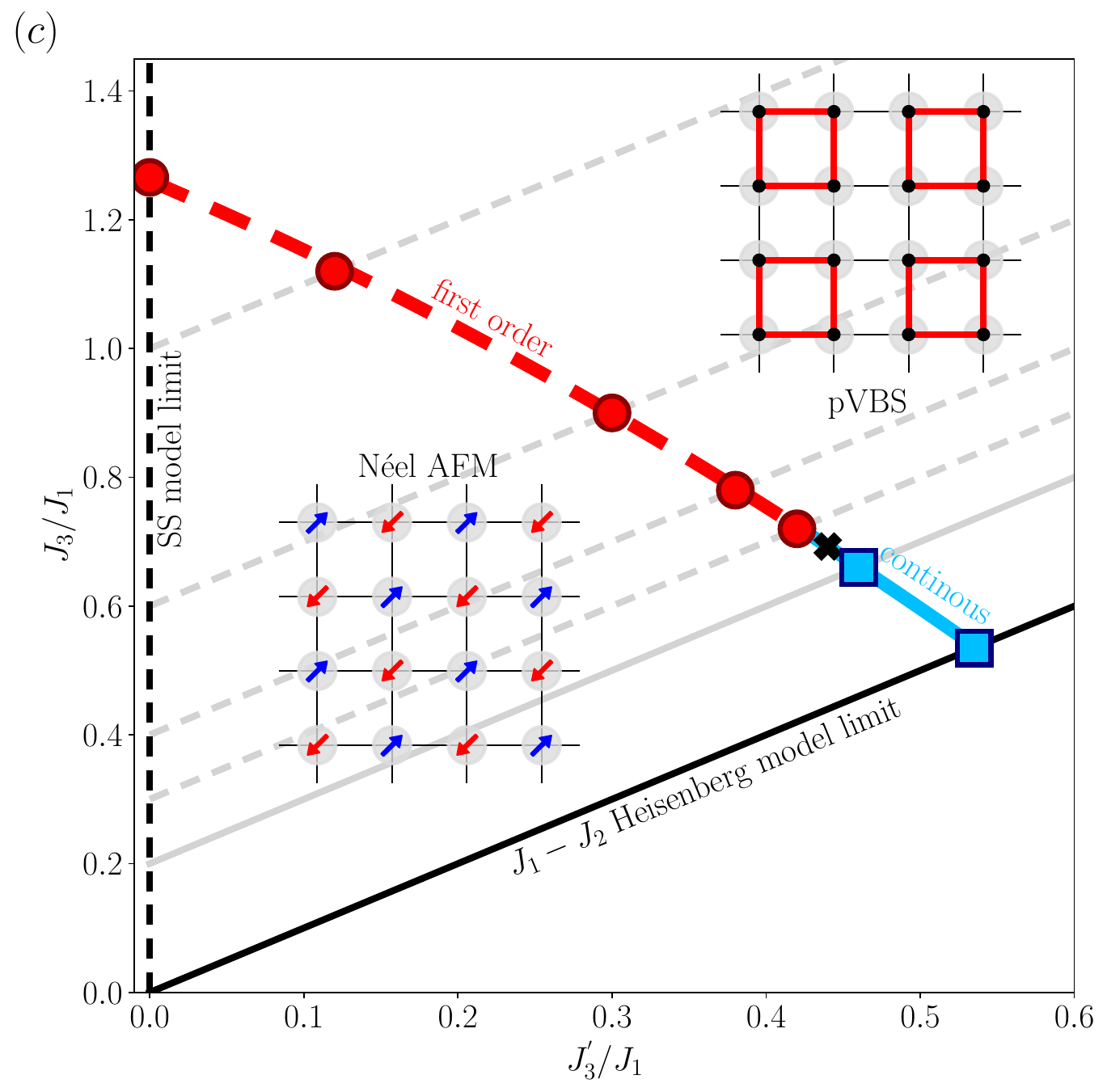}
       \caption{{\bf Model and phase diagram.} (a) The illustration of the model we propose to bridge the Shastry-Sutherland model (SS model) and the $J_1$-$J_2$ Heisenberg model. The interactions $J_1$, $J_3^{\prime}$ and $J_3$ are represented by solid black, dashed red, and solid blue lines, respectively. (b) The ground-state phase diagram of the pure Shastry-Sutherland model ($J_3^{\prime} = 0$), with $\alpha=J_1/J_3$, featuring three distinct phases: the dimer valence bond state (dVBS), the plaquette valence bond state (pVBS), and the N\'eel antiferromagnetic (AFM) phase. The transition from the dVBS to the pVBS is well known to be a strongly first-order transition. In this work, we find a direct weak first-order phase transition between the pVBS and the N\'eel AFM phases. (c) The phase diagram of the generalized Shastry-Sutherland model in (a), where the dashed red lines indicate first-order transitions and the solid blue lines represent continuous transitions between the pVBS and N\'eel AFM phases. We identify an exotic tri-critical point (indicated by the black crossing symbol) along the transition line from the pVBS to the AFM phase, where the transition changes from first-order in the Shastry-Sutherland limit to continuous in the $J_1$-$J_2$ Heisenberg limit.
       }
       \label{Phase_diagram}
\end{figure*}

The Shastry-Sutherland model is proposed to describe the magnetic material $\text{SrCu}_2(\text{BO}_3)_2$ \cite{PhysRevLett.82.3168,Jim_nez_2021,doi:10.1126/science.adc9487}, with the effective spin Hamiltonian
\begin{equation} 
	H = J_1\sum_{\langle i,j \rangle}S_{i} \cdot S_{j}+J_3\sum_{\langle\langle i,j \rangle\rangle \in \textbf{\color{deepskyblue} blue}}S_{i} \cdot S_{j}
\end{equation}
where $S_i$ is the spin-1/2 operator on site $i$, and the summations are taken over nearest-neighbor black pairs ($\langle i,j \rangle$) and part of the next-nearest-neighbor ($\langle \langle i,k \rangle \rangle$) blue pairs as shown in Fig.~\ref{Phase_diagram}(a). The SS model is renowned for its potential to host a deconfined quantum critical point (DQCP)~\cite{doi:10.1126/science.1091806, PhysRevB.70.144407,senthil2023deconfined,PhysRevB.98.085140,PhysRevX.9.041037}, a continuous phase transition between two distinct symmetry-breaking phases, which is beyond the conventional Landau-Ginzburg paradigm, between the plaquette valence bond state (pVBS) and the N\'eel antiferromagnetic (AFM) phases. The SS model contains rich phases.  As the ratio $J_1/J_3$ (defined as $\alpha$ in this work) increases, the ground state undergoes a series of quantum phase transitions, beginning with a dimer singlet phase, transitioning to a plaquette valence bond state (VBS) at approximately $J_1/J_3=0.675$, and ultimately evolving into a N\'eel antiferromagnetic (AFM) state at around $J_1/J_3=0.79$~\cite{PhysRevB.105.L060409,PhysRevB.107.L220408,PhysRevLett.133.026502,PhysRevB.87.115144}. While the transition from the dimer phase to the plaquette VBS phase is distinctly first-order, the characteristics of the transition from the plaquette VBS to the AFM phase remain ambiguous, with various studies suggesting different possibilities: a continuous phase
transition~\cite{PhysRevX.9.041037,PhysRevLett.131.116702,PhysRevLett.133.026502}, a first-order transition~\cite{PhysRevB.87.115144,PhysRevB.107.L220408,2024arXiv241100301C}, or the presence of a narrow quantum spin liquid phase between the pVBS
and N\'eel AFM phases~\cite{PhysRevB.105.L060409,2023arXiv231116889L}.

On the other hand, the $J_1$-$J_2$ Heisenberg model is renowned for its potential to host a quantum spin liquid phase, driven by the interplay of nearest and next-nearest neighbor interactions on a square lattice. The Hamiltonian for this model is defined by
\begin{equation}
	H = J_1\sum_{\langle i,j \rangle}S_{i} \cdot S_{j}+J_2\sum_{\langle\langle i,j \rangle\rangle}S_{i} \cdot S_{j}
    \label{Sha_eq}
\end{equation}
where the summations are taken over all the nearest-neighbor ($\langle i,j \rangle$) and next-nearest-neighbor ($\langle \langle i,k \rangle \rangle$) pairs. Recent studies have focused on exploring the non-magnetic regime, which roughly spans the range  $0.5 <J_2/J_1<0.6$. Within this intriguing region, a variety of competing states have been identified using multiple numerical methods. These states include columnar valence bond state (cVBS)~\cite{PhysRevLett.65.1072,PhysRevLett.66.1773,PhysRevB.60.7278}, pVBS~\cite{PhysRevLett.84.3173,PhysRevB.74.144422,PhysRevLett.113.027201,PhysRevB.85.094407}, and potential intermediate quantum spin liquids ~\cite{PhysRevB.88.060402,PhysRevLett.121.107202,PhysRevX.11.031034,LIU20221034,PhysRevB.86.024424}. Notably, recent calculations have reported that there is a direct phase transition between the pVBS~\cite{PhysRevB.110.195111}\footnote{In the $J_1$-$J_2$ Heisenberg model, a rigorous analysis to pin down plaquette over columnar VBS has been performed in Ref.~\cite{PhysRevB.110.195111} using an infinitesimal symmetry breaking field and a proper finite size scaling.} and the Neel phases at $J_2/J_1 = 0.535(3)$~\cite{PhysRevB.109.L161103}, indicating the absence of intermediate quantum spin liquid phase in this model~\cite{PhysRevB.109.L161103,PhysRevB.109.235133}.

{\em Model and Methods--}
In this study, we propose a model that bridges the SS model and the $J_1$-$J_2$ Heisenberg model, aiming to unify the understanding of these two foundational models. 
Notably, prior numerical results have indicated evidence of a continuous transition in the $J_1$-$J_2$ model compared to the Shastry-Sutherland model. By tracking the evolution of this exotic phase within the phase diagram of our unified model, we aim to uncover new perspectives on quantum criticality and the emergent phenomena arising from the interplay of frustrating interactions. 
The proposed model is described by the Hamiltonian
\begin{equation}  
	H = J_1\sum_{\langle i,j \rangle}S_{i} \cdot S_{j}+J_3 \hspace{-13pt} \sum_{\langle\langle i,j \rangle\rangle \in \textbf{\color{deepskyblue} blue}} \hspace{-13pt} S_{i} \cdot S_{j} +J_3^{\prime} \hspace{-10pt} \sum_{\langle\langle i,j \rangle\rangle \in \textbf{\color{red} red}} \hspace{-11pt}  S_{i} \cdot S_{j}
    \label{model}
\end{equation}
and the summations are taken over nearest-neighbor ($\langle i,j \rangle$) and next-nearest-neighbor ($\langle \langle i,j \rangle \rangle$) blue and red pairs as shown in Fig.~\ref{Phase_diagram}(a). This model encompasses the SS model and the $J_1$-$J_2$ Heisenberg in the limits 
$J^{\prime}_3=0$ and $J^{\prime}_3=J_3$, respectively.

Recent advancements in the density matrix renormalization group (DMRG) \cite{PhysRevLett.69.2863} have introduced a new method known as the fully augmented matrix product state (FAMPS) \cite{Qian_2023}, which significantly enhances our ability 
to accurately determine the ground state of large systems by incorporating an additional disentangler layer on matrix product state (MPS) \cite{SCHOLLWOCK201196}. This method enables calculations on cylinders (which are periodic in the y-direction and open in the x-direction as illustrated in Fig.~\ref{Phase_diagram}(a)) with circumferences up to $L=16$. Unless otherwise specified, we concentrate on $L\times 2L$ cylinder systems following \cite{PhysRevB.105.L060409}. 
For $L\leq 14$, we present simulation results without disentanglers, i.e., employing the regular SU(2)-symmetric DMRG, and we push the bond dimension $D$ up to $15000$ SU(2) multiples, equivalent to about $60000$ U(1) states to ensure high simulation accuracy. 

For $L=16$, we employ the FAMPS method and push the bond dimension to as large as $D=20000$ U(1) states. {Our test indicates that FAMPS results with bond dimension $D$ have almost the same accuracy as the MPS results with bond dimension $4D$ for the $L=16$ results. FAMPS results also make the extrapolation with truncation error more reliable \cite{PhysRevB.110.195111}.} This setup ensures that all truncation errors in our calculations are below or at approximately $1\times 10^{-5}$. The $L = 16$ calculation takes about three weeks on three $64$-core nodes for one parameter. All results presented in this work have been extrapolated to zero truncation errors. The extrapolation processes are shown in the supplementary materials.


The boundary of the VBS phase can be accurately identified by examining the energy crossing between the first excited states in the $S\,{=}\,1$ and $S\,{=}\,0$ sectors~\cite{PhysRevLett.121.107202,K_Nomura_1995,OKAMOTO1992433,PhysRevLett.115.080601,10.1063/1.3518900,PhysRevB.94.144416,PhysRevLett.121.107202, PhysRevX.11.031034,PhysRevB.105.L060409,PhysRevB.109.L161103}, which shows small finite size effect \cite{OKAMOTO1992433,PhysRevB.105.L060409}. In contrast, the boundary of the N\'eel AFM phase is determined by analyzing the crossing of the correlation ratio of the AFM order parameter~\footnote{$m_s^2(\bm{k})=\frac{1}{L^4} \sum_{\bm{ij}}\langle S_{\bm{i}}\cdot S_{\bm{j}} \rangle e^{i\bm{k}(\bm{i}-\bm{j})}$, $\bm{i}=(i_x,i_y)$ is the site position and the summation runs over the center $L\times L$ sites.} 
\begin{align}
    \xi_m/L=\frac{1}{2\pi}\sqrt{\frac{m_s^2(\pi,\pi)}{m_s^2(\pi+2\pi/L,\pi)}-1},
\end{align}
which shows smaller finite-size corrections than other quantities~\cite{10.1063/1.3518900,PhysRevLett.115.157202,PhysRevB.109.L161103}. We also show the results of $m_s^2(\pi,\pi)$ in the supplementary materials. We find that it is difficult to infer the precise location of the boundary for AFM phases from $m_s^2(\pi,\pi)$, because of the large uncertainty in the extrapolation of $m_s^2(\pi,\pi)$ to the thermodynamic limit when the studied sizes are not large enough.

To differentiate between first-order and continuous transitions, we utilize the first derivative of the ground state energy. For a first-order transition, finite-size scaling analysis of the first derivative of ground state energy reveals a crossing point, with a deviation from the transition point in the thermodynamic limit bounded by $O(e^{-L})$~\cite{1990JSP6179B}. 
In our simulation, the derivative of energy density with the parameters in the studied model corresponds to the expectation value of the following operators by the Feynman-Hellmann theorem: $\frac{1}{N}\sum_{\langle\langle i,j \rangle\rangle}  \langle  S_{i} \cdot S_{j} \rangle$ for the interpolated model and $\frac{1}{N} \sum_{ \langle i,j \rangle }  \langle  S_{i} \cdot S_{j} \rangle$ for the pure Shastry-Sutherland model, because in the Shastry-Sutherland model $J_3$ is set as energy unit, while in the generalized model case, $J_1$ is set as energy unit.
Note that these operators differ from the energy density, which should be continuous across the transition regardless of whether it is first or second order. This quantity can exhibit the crossing behavior only if it takes different values for the VBS and AFM phases, implying that the transition is discontinuous with phase coexistence.
In contrast, such a crossing point is absent in continuous transitions. 
Indeed, this crossing point has been numerically observed in previous calculations~\cite{PhysRevB.109.L161103}, allowing for a precise determination of the first-order transition point.  

\begin{figure}[t]
    \includegraphics[width=40mm]{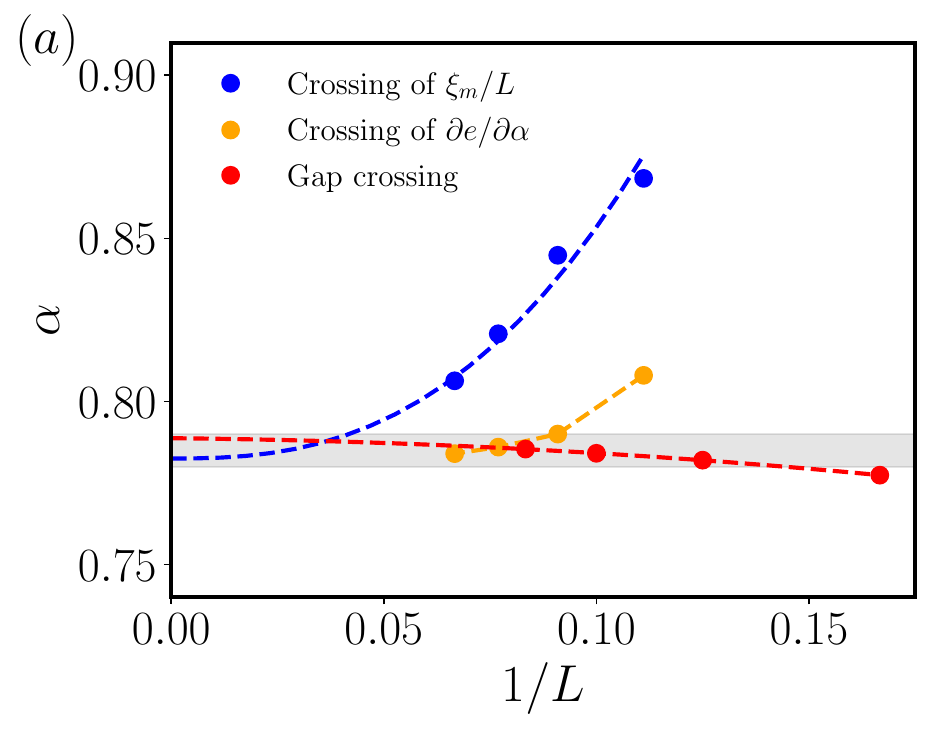}
    \includegraphics[width=40mm]{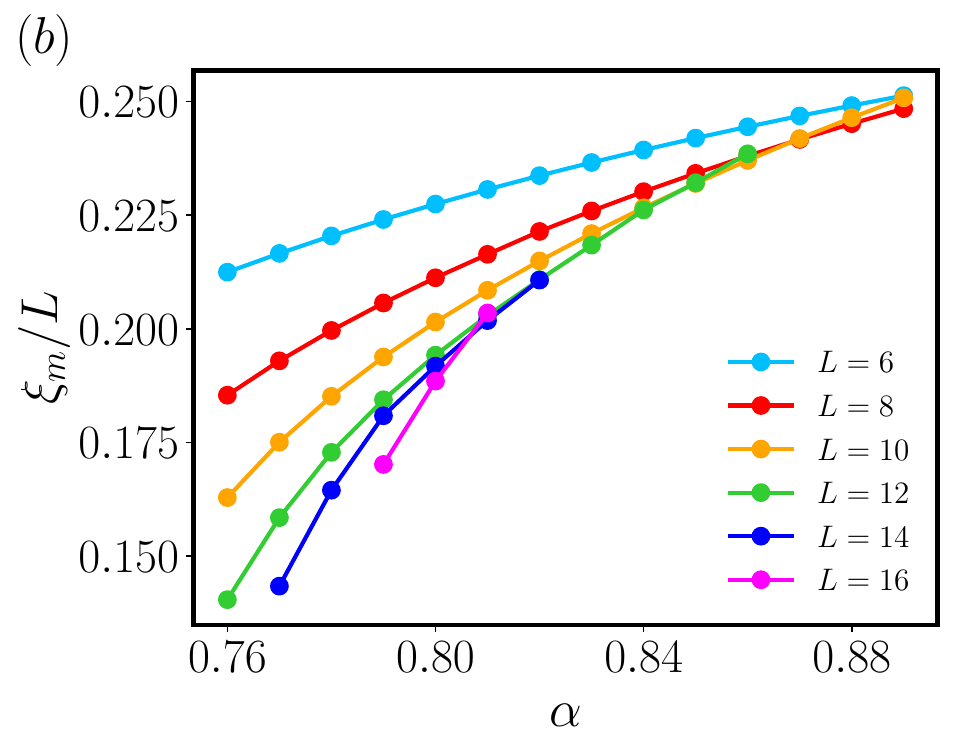}
    \includegraphics[width=40mm]{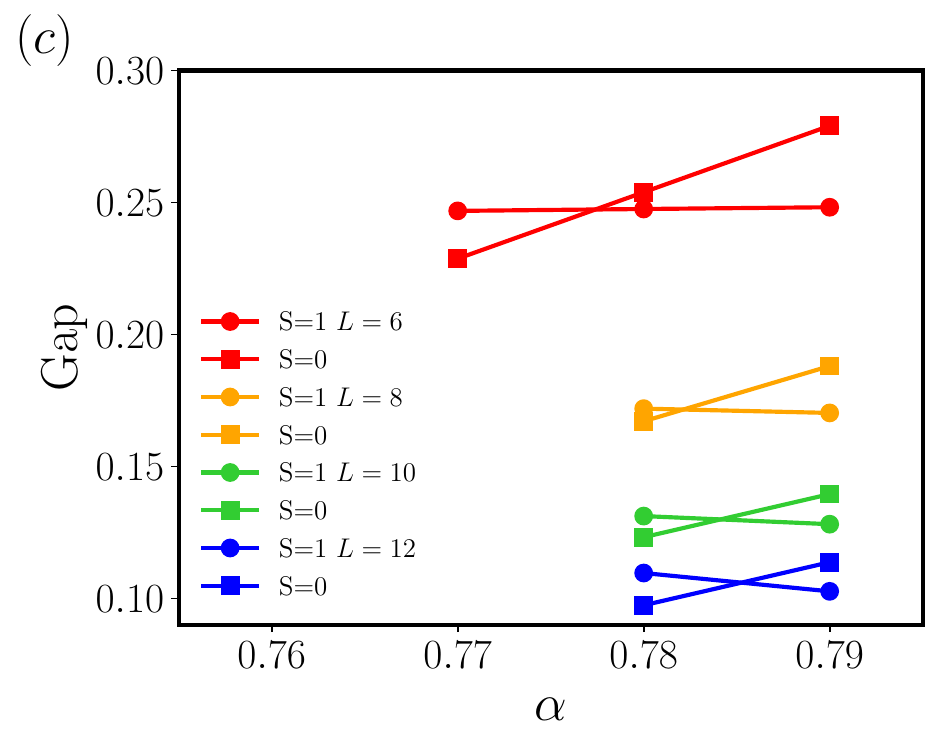}
    \includegraphics[width=40.5mm]{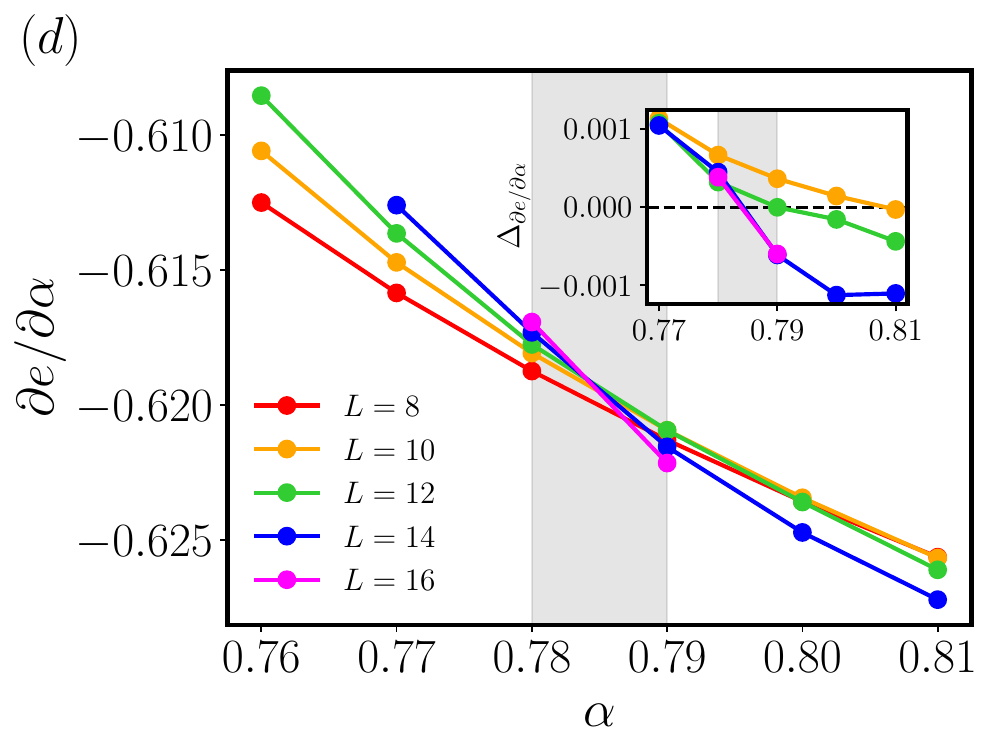}
       \caption{{\bf Phase boundary in the SS model.} {\bf (a)} The results for the finite-size crossing points derived from the AFM correlation ratio, singlet and triplet gap crossing points, and the crossing points of the first-order derivative of the ground state energy with respect to $\alpha$, which suggest a first-order transition at $\alpha=J_1/J_3=0.785(5)$ (shown as the gray band in (a)). Here, the crossing points between $L$ and $L+2$ from (b,d) are plotted with respect to the middle value $L\,{+}\,1$. {\bf(b)} The results of the AFM correlation ratio $\xi_m/L$.  
       {\bf(c)} Singlet and triplet gaps versus $\alpha$ in the neighborhood of the expected quantum phase transition point. {\bf (d)} The crossing points for the first derivative of ground state energy with respect to $\alpha$ calculated using the Feynman-Hellmann theorem. This suggests a first-order transition at $\alpha=0.785(5)$ as shown by the gray band. (Inset) The difference of $\partial e/\partial \alpha$ between $L$ and $L-2$.
       }
       \label{Shastry_Sutherland}
\end{figure}

\begin{figure}[t]
    \includegraphics[width=70mm]{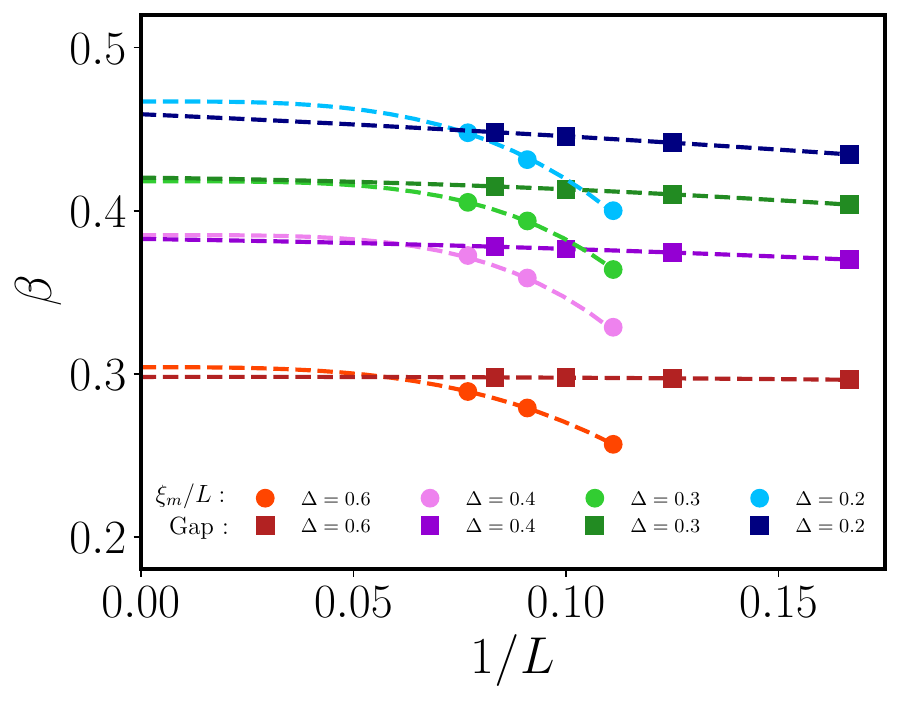}
       \caption{ {\bf Phase boundary in the generalized model.}
       The crossing points extracted from the AFM correlation ratio $\xi_m/L$ and the lowest singlet and triplet gaps for different $\Delta=J_3-J_3^{\prime}$ values. While the correlation ratio data is obtained for up to $L\,{=}\,14$, the singlet-triplet crossing point is obtained for system up to $L\,{=}\,12$ as excited states require more computational resources and its finite size effect is small.
       Our results indicate that as the system evolves from the SS model limit to the $J_1$-$J_2$ Heisenberg limit, it always experiences a direct transition between the N\'eel AFM and pVBS phases.
       }
       \label{Crossing_all}
\end{figure}

\begin{figure}[t]
    \includegraphics[width=42mm]{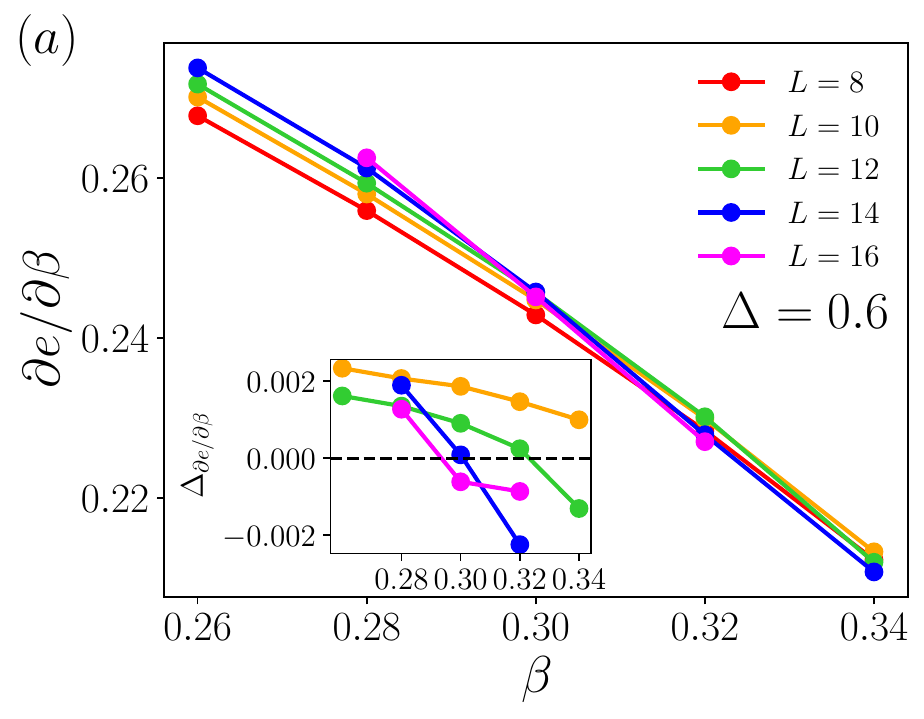}
    \includegraphics[width=42mm]{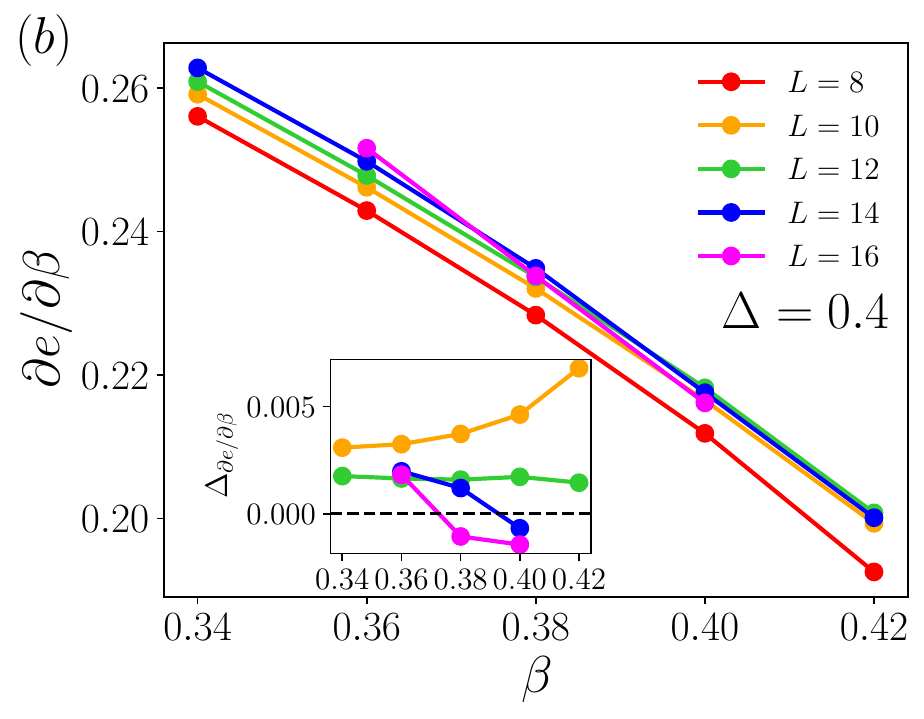}
    \includegraphics[width=42mm]{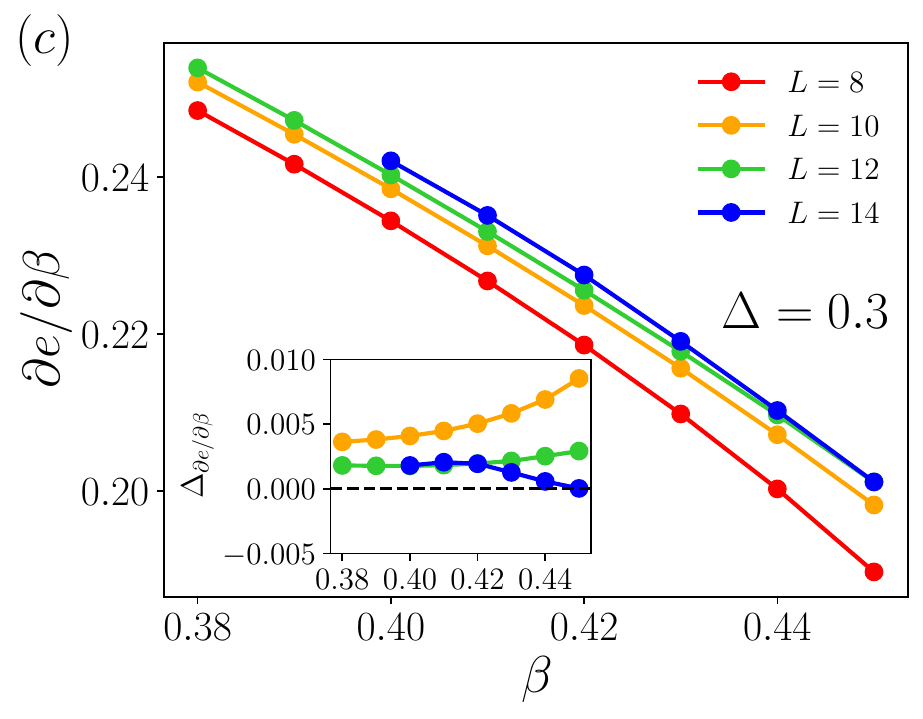}
    \includegraphics[width=42mm]{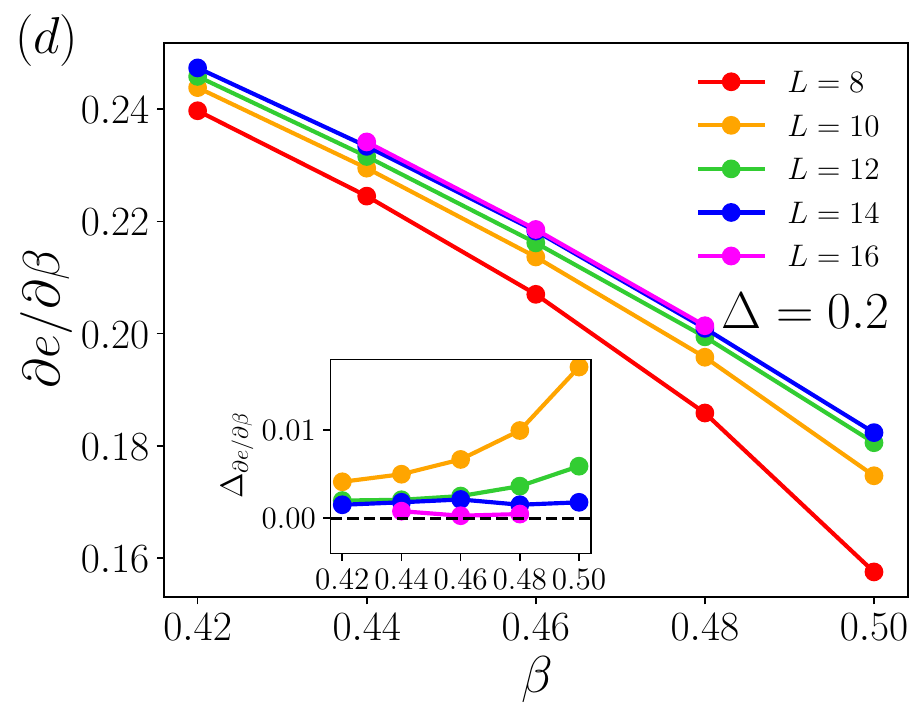}
       \caption{ {\bf Energy derivative.} The results of the first-order derivative of ground state energy as a function of $\beta=J_3^{\prime}/J_1$ for different $\Delta=J_3-J_3^{\prime}$ values at the neighborhood of the transition point shown in Fig.~\ref{Crossing_all}. 
       In the inset we plot the difference of the first order derivative between $L$ and $L-2$ systems, to show the crossing points clearer. The
       crossing point for the first derivative of the ground state
       energy disappears as the system approaches the $J_1$-$J_2$
       Heisenberg model limit. This result suggests that the transition between the N\'eel AFM and pVBS phases ultimately turns to continuous when the model is close to the $J_1$-$J_2$ Heisenberg model limit~\cite{PhysRevB.109.L161103}. From the results we can infer that the change from first order to continuous transition occurs at approximately between $\Delta = 0.2$ and $\Delta = 0.3$ which is marked as the black crossing symbol in Fig.~\ref{Phase_diagram}. }
       \label{Crossing_dE}
\end{figure}

{\em Results for the pure Shastry-Sutherland Model --}
First, we study the pure Shastry-Sutherland model with $J_3^{\prime}=0$. Its phase diagram is depicted in Fig.~\ref{Phase_diagram}(b). We explore the phase diagram by varying $\alpha=J_1/J_3$. The transition between the dimer valence bond state (dVBS) and plaquette valence bond state (pVBS) phases is well known to be a first-order one \cite{PhysRevB.87.115144,PhysRevB.107.L220408}. However, the transition between the pVBS and N\'eel antiferromagnetic (AFM) phases has been the subject of extensive research, with various studies suggesting different possibilities: a continuous phase transition, a first-order transition, or the presence of a narrow quantum spin liquid phase between the pVBS and N\'eel AFM phases.

Our main results for the SS model are shown in Fig.~\ref{Shastry_Sutherland}, with a summary in Fig.~\ref{Shastry_Sutherland}(a). In Fig.~\ref{Shastry_Sutherland}, we can find a touch between the extrapolation results from the correlation ratio crossings for the boundary of the N\'eel AFM phase and the $S\,{=}\,0$ and $S\,{=}\,1$ gap crossing for the boundary of the pVBS phase, demonstrating a direct transition between the N\'eel AFM and pVBS phases at $\alpha=0.785(5)$. 
The data used to extract the crossing point are shown in Fig.~\ref{Shastry_Sutherland}(b) and Fig.~\ref{Shastry_Sutherland}(c) for the correlation ratio and gap crossing, respectively. 
The results of VBS order parameters $D_a=\frac{1}{N_b}\sum_r e^{i q_a \cdot r} S_r \cdot S_{r+a}$ ($a = x/y$)~\footnote{$N_b$ is the number of bonds in the calculation, $q_a$ is the wave vector associated with the VBS order, taking values $(\pi, 0)$ or $(0, \pi)$ depending on the direction $a$ of the dimer order. The summation runs over the center $L\times L$ sites.} can be found in the supplementary materials, which also suggests a transition point between $\alpha = 0.78$ and $\alpha = 0.80$ and the VBS are pVBS type. It is recently shown that the VBS state in the $J_1$-$J_2$ Heisenberg model is pVBS type \cite{PhysRevB.110.195111}. So there is only one PVBS phase in the phase diagram as shown in Fig.~\ref{Phase_diagram}.
Additionally, we compute the first-order derivative of the ground state energy with respect to $\alpha$ using the Feynman-Hellmann theorem, as shown in Fig.\ref{Shastry_Sutherland} (d) \footnote{Here, we show the results for $L\times 2.5L$ systems, which reveal a more pronounced crossing point than $L\times 2L$ systems. Results for $L\times 2L$ systems, provided in the supplementary material, also exhibit a crossing point at the transition.}.  
The first-order derivative of the ground state energy confirms a crossing point at the transition point $\alpha_c=0.785(5)$ which agrees with the transition point determined from the crossing points of correlation ratio and singlet and triplet gap in Fig.~\ref{Shastry_Sutherland}, demonstrating that the transition is a first-order one in this model.

{\em Bridging the Shastry-Sutherland Model to the $J_1$-$J_2$ Heisenberg Model--} 
In this section, we explore how the transition point between the pVBS and N\'eel AFM phases evolves as the system shifts from the SS limit to the $J_1$-$J_2$ Heisenberg limit. A key and intriguing question we address is how this transition progresses from the first-order regime: will it become continuous, as anticipated for a DQCP, or will it lead to the emergence of a spin-liquid phase? 

To answer this question, we set the parameter in Eq.~(\ref{model}) as $ J_3 = J_3^{\prime} + \Delta $ and investigate the phase diagram by varying $ \beta = J_3^{\prime}/J_1$, with $J_1$ serving as the energy unit (different to the convention used in the pure Shastry-Sutherland model, where $\alpha$ is defined as $J_1/J_3$ and $J_3$ is set as energy unit). We explore the phase diagram for values of $\Delta = \{1.0, 0.6, 0.4, 0.3, 0.2\}$, represented by the gray lines in Fig.~\ref{Phase_diagram}(c). 
Our results indicate the model exhibits a direct transition between the N\'eel AFM and pVBS phases through the range of $\Delta$ we studied. In Fig.~\ref{Crossing_all}, the correlation ratio and gap crossing analyses indicate that the boundaries of the AFM and VBS phases touch with each other for all examined values of $\Delta$, although numerically we cannot rule out a possible very narrow intermediate region. 
Importantly, the first derivative of the ground state energy, presented in Fig.~\ref{Crossing_dE}, reveals that the crossing point for the first derivative of the ground state energy disappears as the system approaches the $J_1$-$J_2$ Heisenberg model limit~\cite{PhysRevB.109.L161103}. 
This suggests that the transition between the N\'eel AFM and pVBS phases ultimately becomes continuous when the model is sufficiently close to the $J_1$-$J_2$ Heisenberg model limit. A summary of the results is shown in Fig.~\ref{Phase_diagram}(c), where the phase boundary is marked based on the linear extrapolation of singlet-triplet crossing points.

The observation that the transition evolves from first-order to second-order suggests the existence of a multicritical point, indicated by the black cross in Fig.\ref{Phase_diagram}. Notably, this phase diagram is consistent with one of the hypothetical scenarios proposed in Ref.~\cite{takahashi2024so5multicriticalitytwodimensionalquantum}, where the sign-problem-free $J$-$Q$ model was studied along a specific line cut exhibiting a first-order transition~\cite{PhysRevResearch.2.033459, So4Bowen}. Our comprehensive numerical investigation using a concrete model raises intriguing questions about the field-theoretic description of this transition. While the conventional theory of deconfined quantum critical points (DQCP)~\cite{senthil2023deconfined} for continuous transitions is well-developed, the theory surrounding tricritical DQCPs remains largely unexplored. A more detailed investigation of this multicritical point is left for future work.

{\em Conclusion and Perspective--}
In this work, we have introduced a generalized Shastry-Sutherland model, which encompasses both the pure Shastry-Sutherland model and the $J_1$-$J_2$ Heisenberg model as special cases within its phase diagram. Through extensive large-scale DMRG calculations and meticulous finite-size scaling, we have determined that the phase transition from the pVBS to the N\'eel AFM phase in the pure Shastry-Sutherland model is a weak first-order transition. Additionally, we have identified a tri-critical point in the phase diagram, bridging the transition from a first-order one in the Shastry-Sutherland limit to a continuous transition in the $J_1$-$J_2$ Heisenberg limit. Our findings highlight the generalized Shastry-Sutherland model as a valuable framework for investigating exotic phases and phase transitions. Furthermore, it can serve as an effective model for describing materials related to the Shastry-Sutherland model. It will also be interesting to study the generalized Shastry-Sutherland model in a magnetic field~\cite{PhysRevLett.112.147203}. 



{\em Acknowledgments--} 
We thank Bryan K. Clark and Yasir Iqbal for fruitful discussions. MQ acknowledges the support from the National Key Research and Development
Program of MOST of China (2022YFA1405400), the National Natural Science Foundation of China (Grant No. 12274290), the Innovation Program for Quantum Science and Technology (2021ZD0301902), and the sponsorship from Yangyang Development Fund. JYL acknowledges the support from the faculty startup grant at University of Illinois, Urbana-Champaign. 

{\em Data availability--} 
The calculation in this work is carried out with TensorKit~\cite{foot7}. The computations in this paper were run on the Siyuan-1 cluster supported by the Center for High Performance Computing at Shanghai Jiao Tong University.

\bibliography{main}

\newpage 

\appendix

\setcounter{equation}{0}
\setcounter{figure}{0}
\setcounter{table}{0}
\makeatletter
\renewcommand{\theequation}{S\arabic{equation}}
\renewcommand{\thefigure}{S\arabic{figure}}
\setcounter{subsection}{0}

\title{Supplementary Materials for ``From the Shastry-Sutherland model to the $J_1$-$J_2$ Heisenberg model''}

\author{Xiangjian Qian}
\affiliation{Key Laboratory of Artificial Structures and Quantum Control (Ministry of Education),  School of Physics and Astronomy, Shanghai Jiao Tong University, Shanghai 200240, China}

\author{Rongyi Lv}
\affiliation{Key Laboratory of Artificial Structures and Quantum Control (Ministry of Education),  School of Physics and Astronomy, Shanghai Jiao Tong University, Shanghai 200240, China}

\author{Jong Yeon Lee} \thanks{jongyeon@illinois.edu}
\affiliation{Department of Physics and Institute of Condensed Matter Theory, University of Illinois at Urbana-Champaign, Urbana, Illinois 61801, USA}

\author{Mingpu Qin} \thanks{qinmingpu@sjtu.edu.cn}
\affiliation{Key Laboratory of Artificial Structures and Quantum Control (Ministry of Education),  School of Physics and Astronomy, Shanghai Jiao Tong University, Shanghai 200240, China}

\affiliation{Hefei National Laboratory, Hefei 230088, China}

\date{\today}

\maketitle

\tableofcontents

\section{Additional results for the Shastry-Sutherland model}

\subsection{Comparison of the ground state energies}
\begin{figure}[!t]
    \includegraphics[width=75mm]{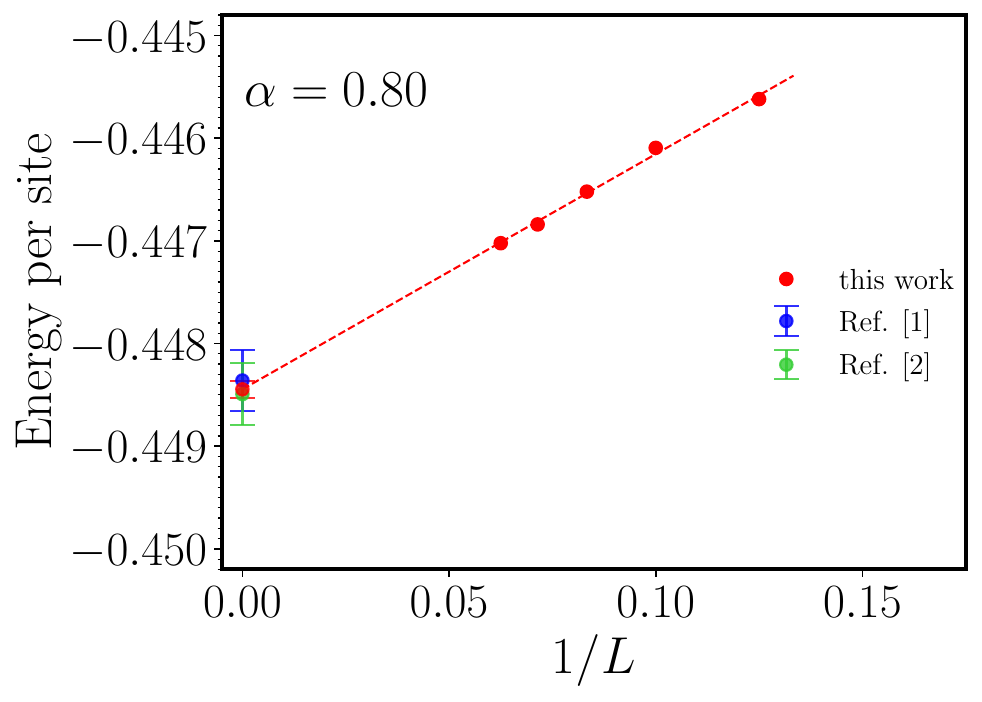}
     \includegraphics[width=75mm]{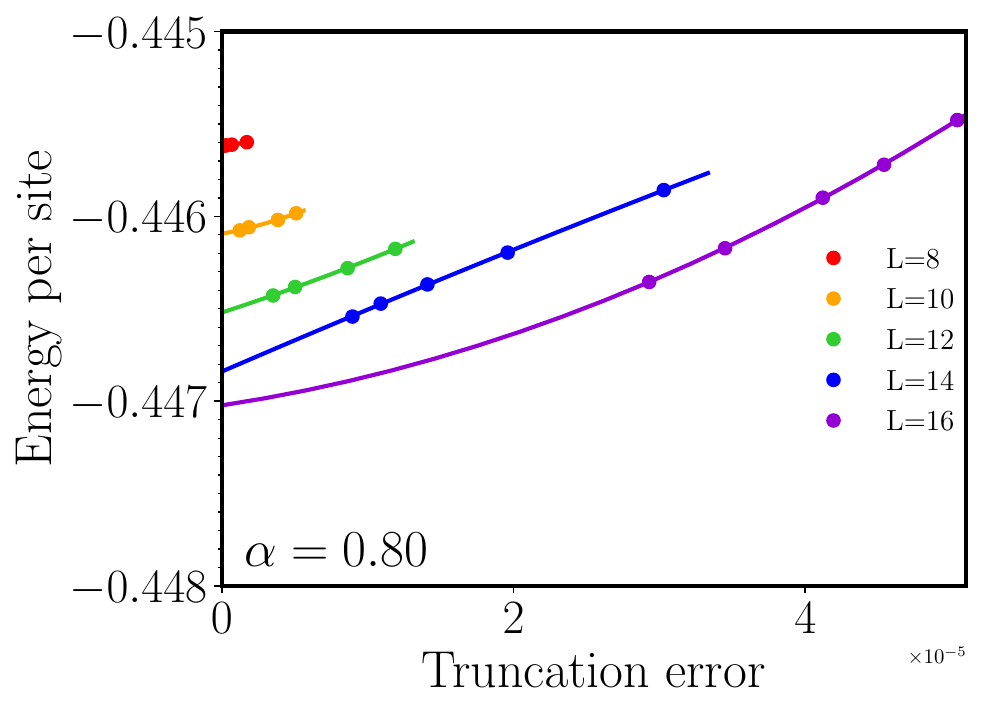}
    \caption{Upper: the extrapolated with truncation error ground state energy per site for $L$ ranging from $8$ to $16$ for the Shastry-Sutherland model with $\alpha = 0.8$ on a $L \times 2L$ cylinders. The extrapolation with truncation error in DMRG/FAMPS can be found in the lower panel. An linear extrapolation of the energy with $1/L$ gives the thermodynamic limit energy as $-0.44844(8)$, consistent with the previous DMRG calculation ($-0.44836(30)$)  \cite{PhysRevB.105.L060409} and a recent calculation on $L\times L$ periodic systems ($-0.44849(30)$) \cite{2023arXiv231116889L}. Lower: the quadratic extrapolation of the energy with truncation error for different sizes.
    }
    \label{Energy_benchmark}
\end{figure}

\begin{figure}[t]
    \includegraphics[width=80mm]{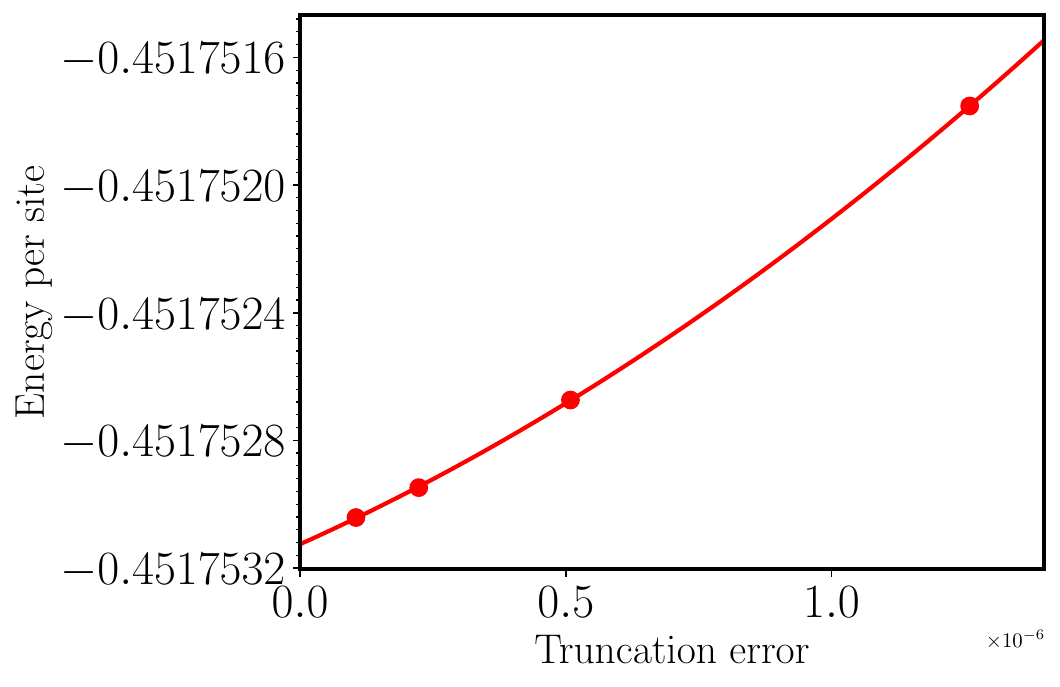}
    \caption{ The ground state energy for a calculated $6\times 6$ PBC system at $\alpha=0.80$ as a function of truncation error. The calculated ground state energy is very close to the exact energy obtained from exact diagonalization $E=-0.4517531$ from Ref.~\cite{2023arXiv231116889L}. The displayed points are the results of calculations performed with bond dimensions of $D=\{30000,40000,50000,60000\}$.}
    \label{Energy_6}
\end{figure}

In Fig.~\ref{Energy_benchmark}, we show the energies of the Shastry-Sutherland model on cylinders with different sizes at $\alpha\,{:=}\,J_1/J_3\,{=}\,0.80$.
The finite size scaling from the linear extrapolation gives energy ($-0.44844(8)$) aligns with a recent extrapolated energy on $L\times L$ periodic systems ($-0.44849(30)$) using neural network method~\cite{2023arXiv231116889L} and the previous DMRG calculation ($-0.44836(30)$)~\cite{PhysRevB.105.L060409}. We also show the ground state energy for a $6\times 6$ PBC system at $\alpha=0.80$ as a function of truncation error in Fig.~\ref{Energy_6}. The extrapolated energy is indistinguishable from the exact diagonalization value $E=-0.4517531$~\cite{2023arXiv231116889L}.

\subsection{The AFM order parameter}
\begin{figure}[t]
    \includegraphics[width=80mm]{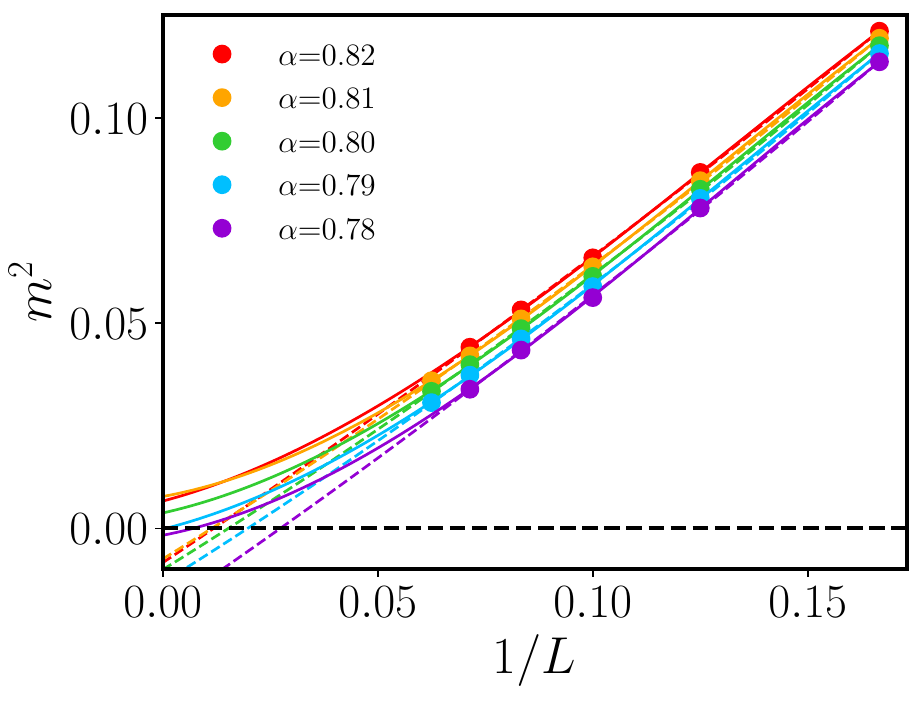}
       \caption{The N\'eel AFM order parameters as a function of circumferences $L$. The solid lines represent a third-order polynomial fit, while the dashed lines indicate a second-order polynomial fit. 
       We can find that the boundary of AFM phase is extremely sensitive to the way we fit the data (first and second-order fit generate unreasonable extrapolation value of negative $m^2$), and this is the reason why we use the correlation ratio to determine the boundary of AFM phase in the main text.  
       }
       \label{AFM_order}
\end{figure}


We calculate the N\'eel AFM order parameter 
\begin{align}
    m_s^2(\bm{k})=\frac{1}{L^4} \sum_{\bm{ij}}\langle S_{\bm{i}}\cdot S_{\bm{j}} \rangle e^{i\bm{k} \cdot (\bm{i}-\bm{j})}
\end{align}
where $\bm{i}=(i_x,i_y)$ and $\bm{k}=(\pi,\pi)$ using the middle region of size $L\times L$ in an open cylinder of size $L \times 2L$. The value of $m^2$ at momentum $(\pi,\pi)$ is shown in Fig.~\ref{AFM_order}.

We also try to determine the AFM phase boundary by extrapolating $m^2$ to the thermodynamic limit. However, we find the extrapolation is highly sensitive to the order of the polynomial fit. 
In particular, first and second-order fit generate unreasonable extrapolation values of negative $m^2$ for $\alpha \in [0.78,0.82]$ and do not yield any discernible transition point. 
Using a third-order polynomial fit, we find a transition point at approximately $\alpha=0.79$, although the use of a third-order polynomial is not fully justified. 
Since the estimation from the squared AFM order parameter does not provide a reliable transition point, we use the correlation ratio to determine the boundary of the AFM phase in the main text.


\subsection{The VBS order parameter}

\begin{figure}[t]
    \includegraphics[width=80mm]{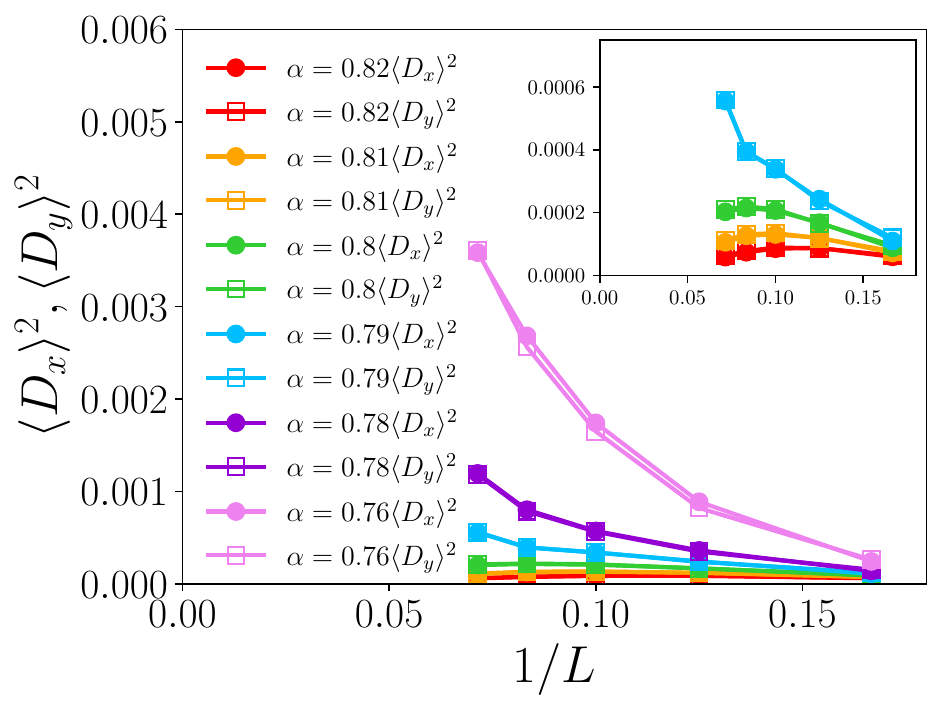}
       \caption{ The VBS order parameters as a function of circumferences $L$. Values of $D_x^2$ and $D_y^2$ are on top of each other. The results suggest a phase transition between $\alpha=0.78$ and $\alpha=0.8$. In the inset, we show a zoom of the results for $\alpha \geq 0.79$. We can find that near the phase transition point, $D_x^2$ and $D_y^2$ behave non-monotonically, which makes the determination of the pVBS boundary difficult from $D_x^2$ and $D_y^2$.   
       }
       \label{VBS_order}
\end{figure}

\begin{figure*}[t]
    \includegraphics[width=170mm]{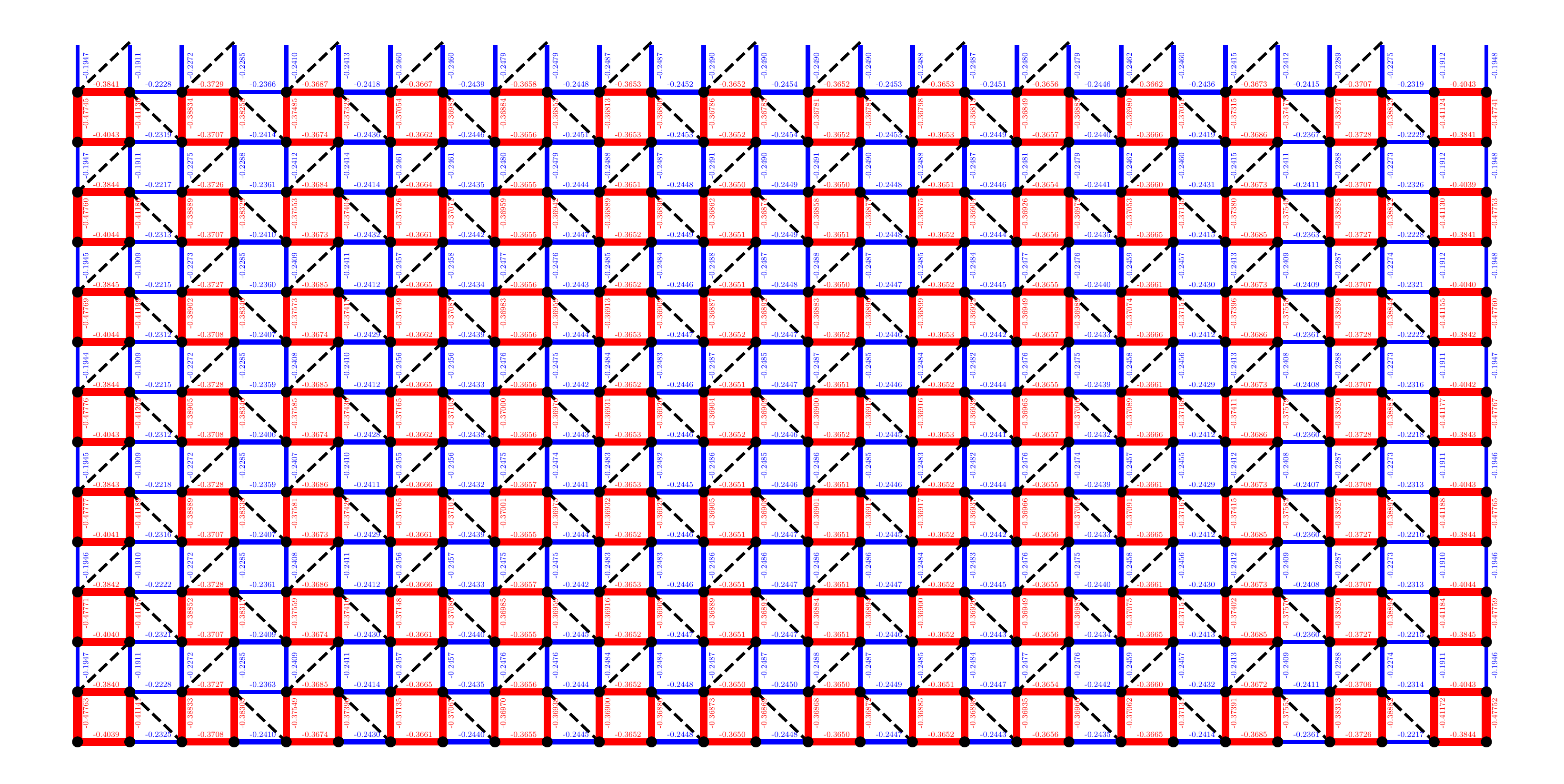}
       \caption{The bond energy, with red/blue indicating lower/higher bond energies, illustrates the spatial structure of the VBS order in the Shastry-Sutherland model for a $14\times 28$ system at $\alpha=0.76$. The plaquettes are formed on the empty squares of the Shastry-Sutherland lattice.}
       \label{VBS_SS}
\end{figure*}

We also calculate the VBS order parameter, as shown in Fig.~\ref{VBS_order}, which shows that $\langle D_x\rangle^2=\langle D_y\rangle^2$, consistent with a plaquette VBS. As depicted in Fig.~\ref{VBS_SS} for $14\times 28$ system at $\alpha=0.76$, plaquettes are formed on the empty squares of the Shastry-Sutherland lattice. The results suggest a phase transition between $\alpha=0.78$ and $\alpha=0.8$.


\subsection{The first derivative of the ground state energy}
\begin{figure}[t]
    \includegraphics[width=80mm]{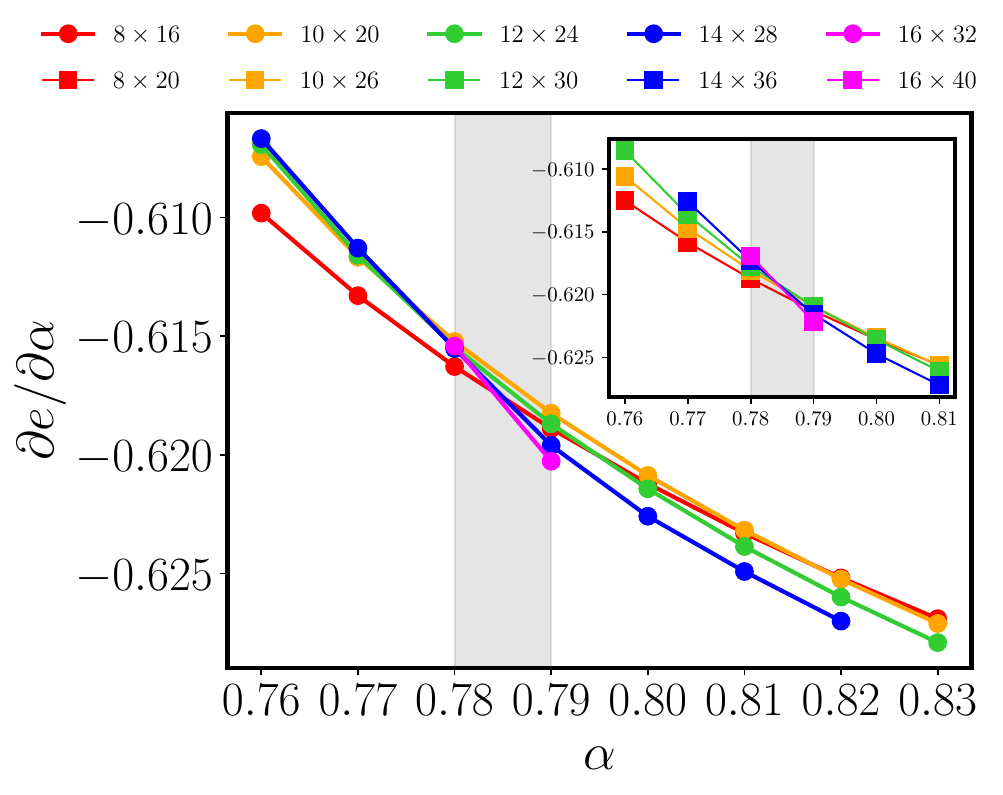}
       \caption{Comparison of the results for the first derivative of the ground state energy calculated by $L\times 2L$ (main plot) and $L\times 2.5L$ systems (inset). Both of the calculated systems demonstrate a crossing point at around $\alpha=0.785(5)$}
       \label{compare}
\end{figure}
As mentioned in the main text, the first derivative of the ground state energy for $L\times2L$ systems also exhibits a crossing point. Here, we compare the results obtained from $L\,{\times}\,2L$ (main plot) and $L\,{\times}\,2.5L$ (inset) systems as illustrated in Fig.~\ref{compare}. Both systems display a crossing point, with the crossing being more pronounced in the $L\times 2.5L$ configurations.

As discussed in the main text, the first derivative corresponds to the expectation value of the following operator: 
\begin{align}
    {\cal O}_1 := \frac{1}{N}\sum_{\langle i,j  \rangle}  \langle  S_{i} \cdot S_{j} \rangle,
\end{align}
which is the average expectation value of the nearest neighbor $S_i \cdot S_j$. The crossing behavior of this operator across the phase transition implies that in the infinite size limit, there would be a discontinuity of the operator. This indicates that the transition would be of a first order.

\section{Generalized model and convergence}

\subsection{The first derivative of the ground state energy}

In the model interpolating the Shastry-Sutherland lattice and $J_1$-$J_2$ model, the first derivative of the ground state energy corresponds to the expectation value of the following operator
\begin{align}
    {\cal O}_2 = \frac{1}{N}\sum_{\langle\langle i,j \rangle\rangle}  \langle  S_{i} \cdot S_{j} \rangle,
\end{align}
which is the average expectation value of the next-nearest neighbor $S_i \cdot S_j$.

\begin{figure*}[t]
    \includegraphics[width=70mm]{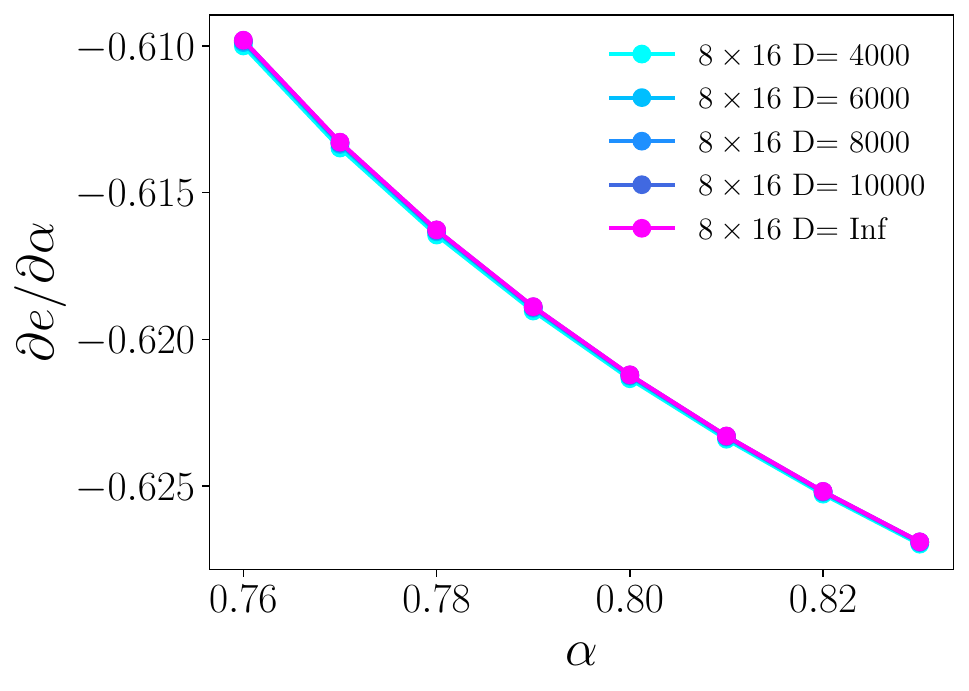}
    \includegraphics[width=70mm]{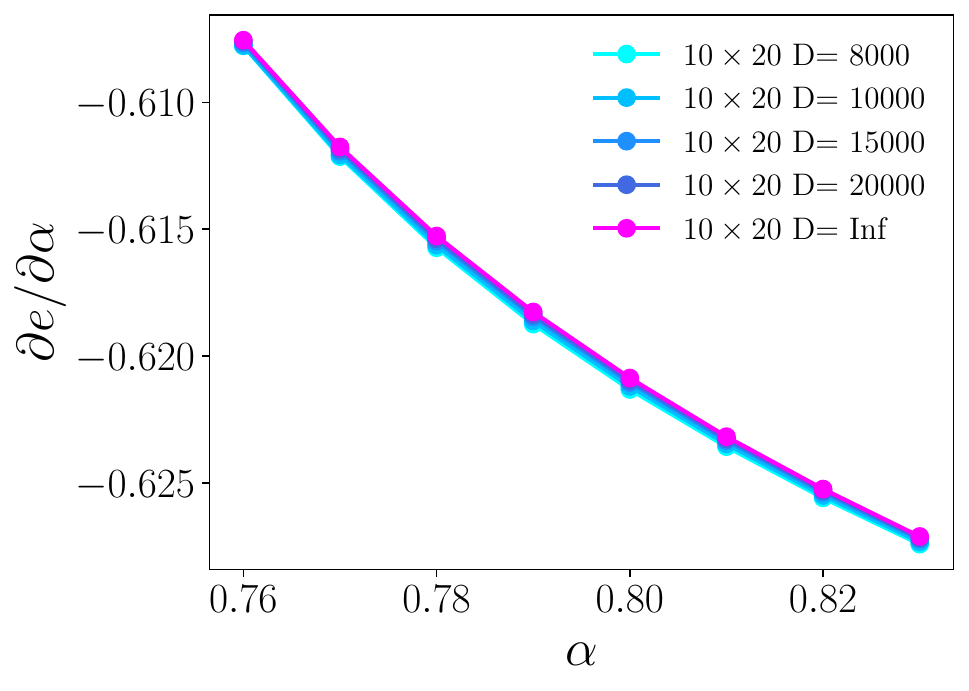}
    \includegraphics[width=70mm]{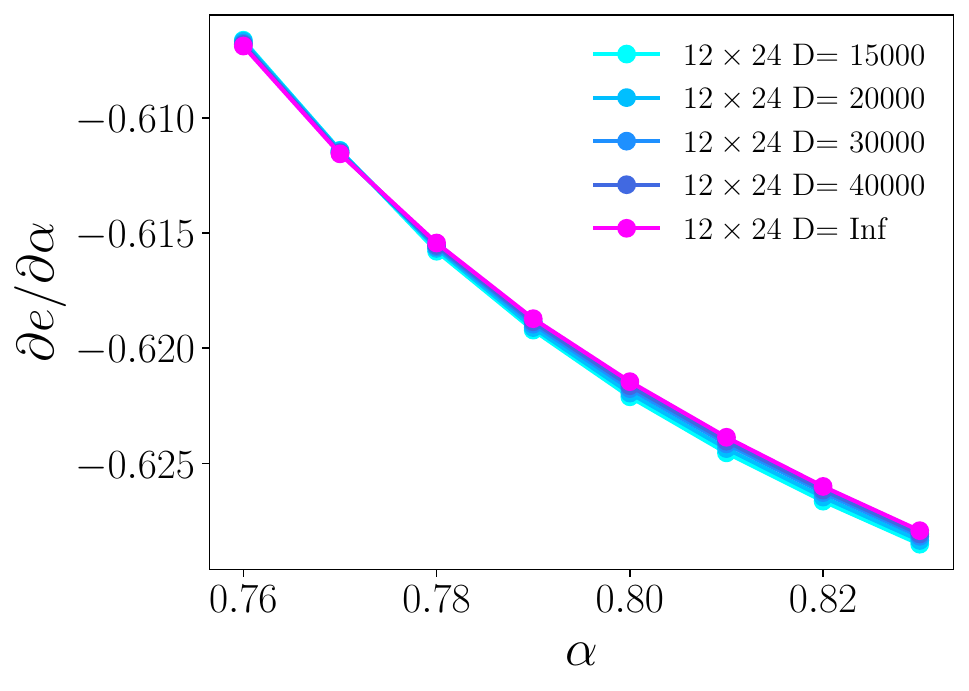}
    \includegraphics[width=70mm]{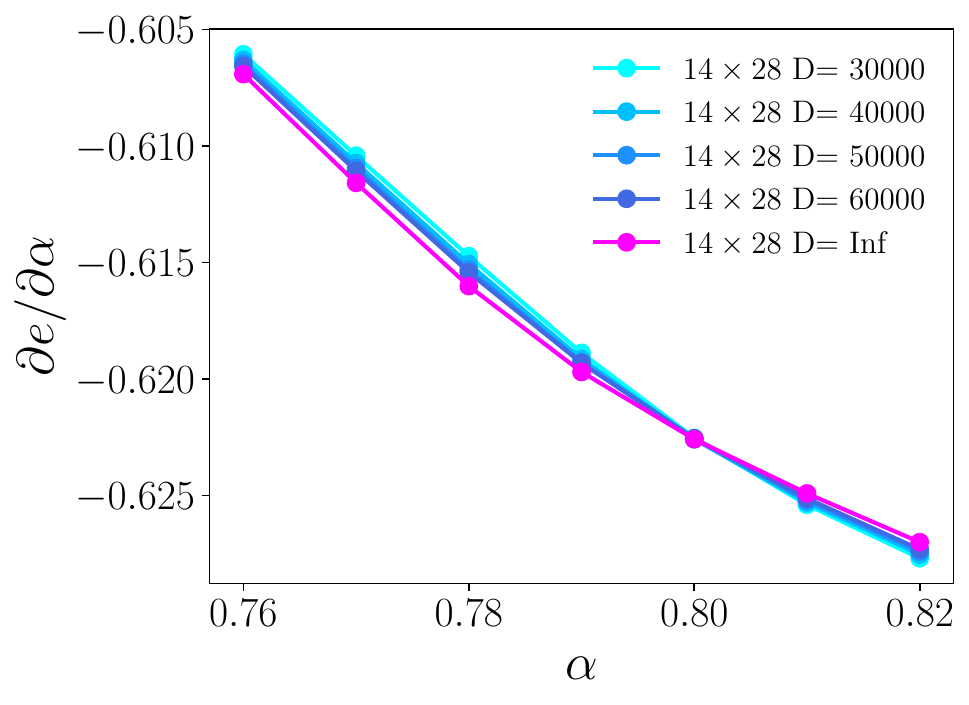}
    \caption{Convergence of the results for the first derivative of the ground state energy for the Shastry-Sutherland model}
    \label{dE_SS}
\end{figure*}

\begin{figure*}[t]
    \includegraphics[width=70mm]{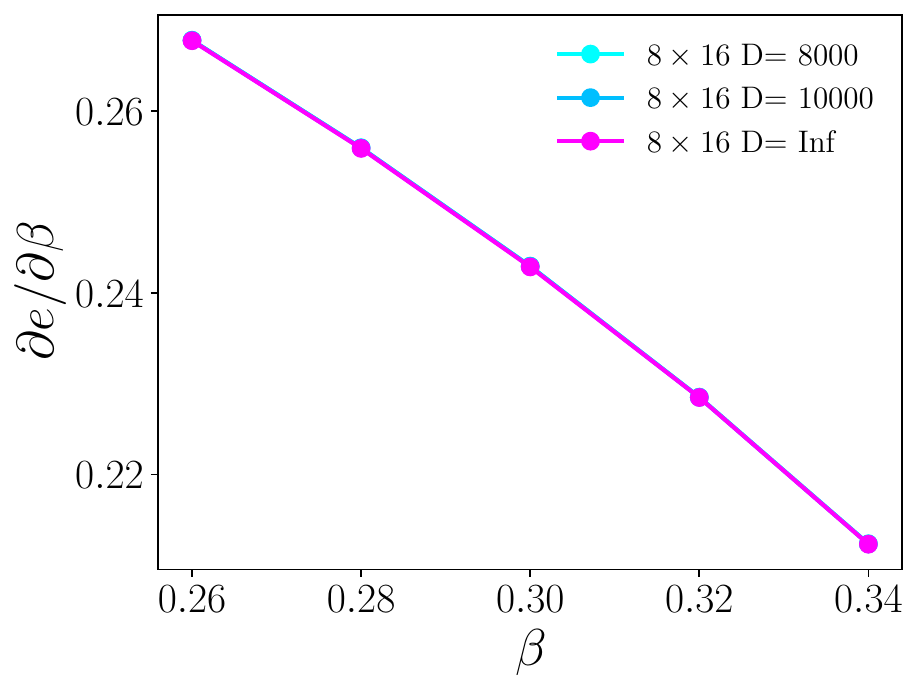}
    \includegraphics[width=70mm]{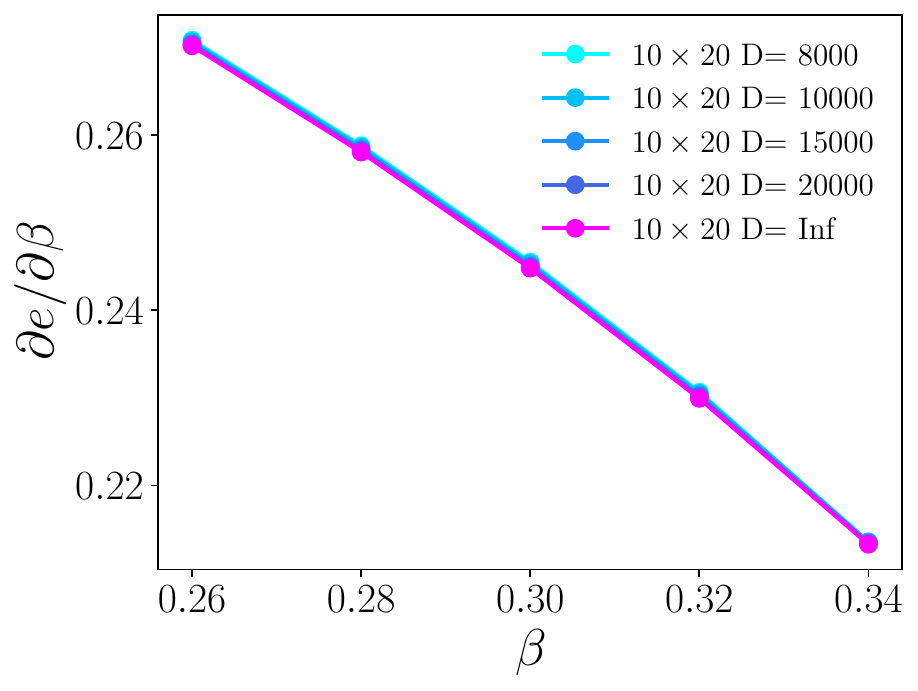}
    \includegraphics[width=70mm]{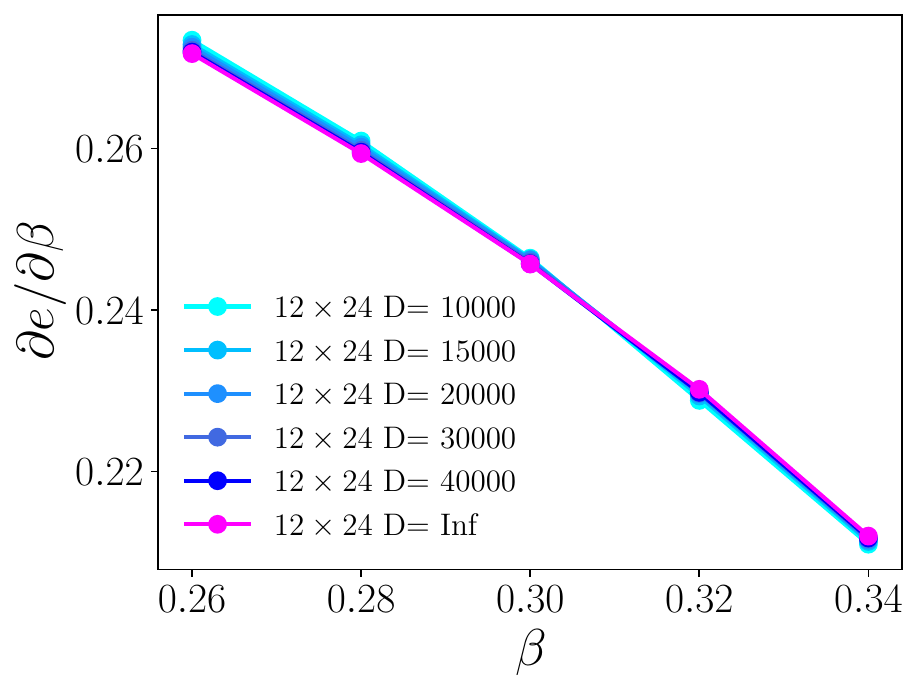}
    \includegraphics[width=70mm]{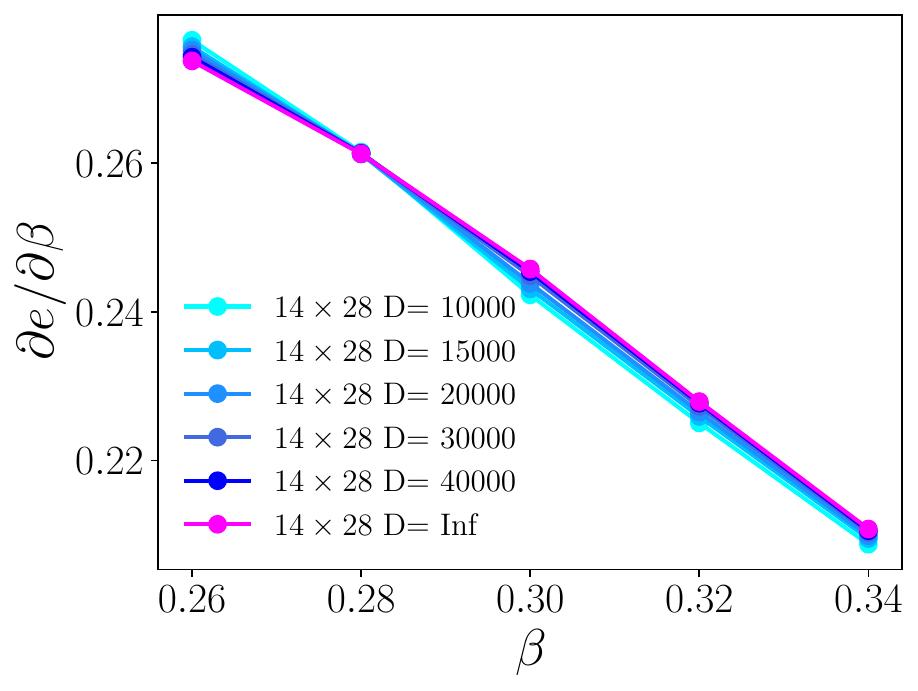}
    \caption{Convergence of the results for the first derivative of the ground state energy for $\Delta=0.6$}
    \label{dE_Delta6}
\end{figure*}

\begin{figure*}[t]
    \includegraphics[width=70mm]{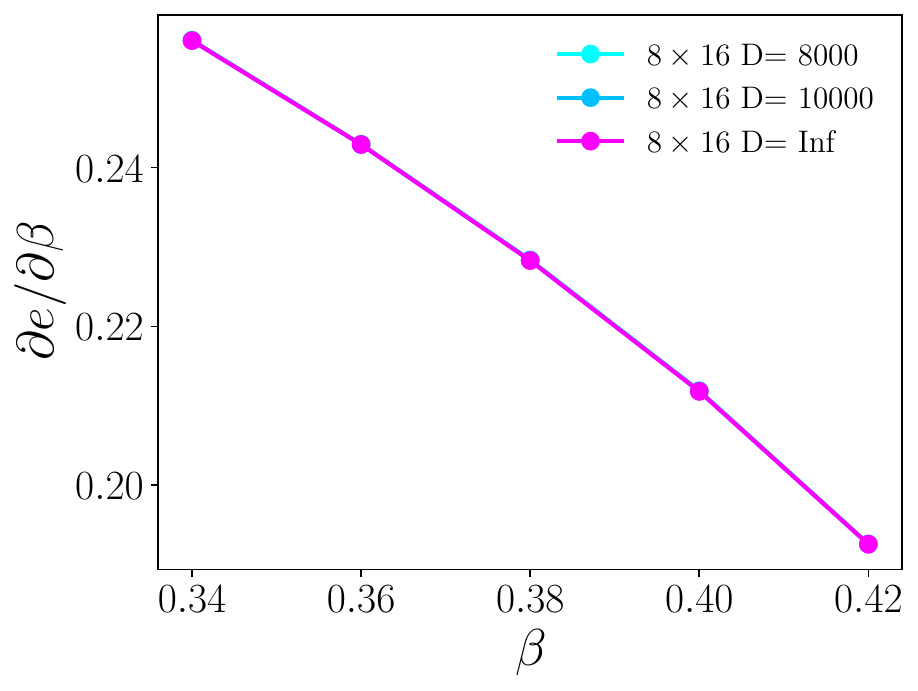}
    \includegraphics[width=70mm]{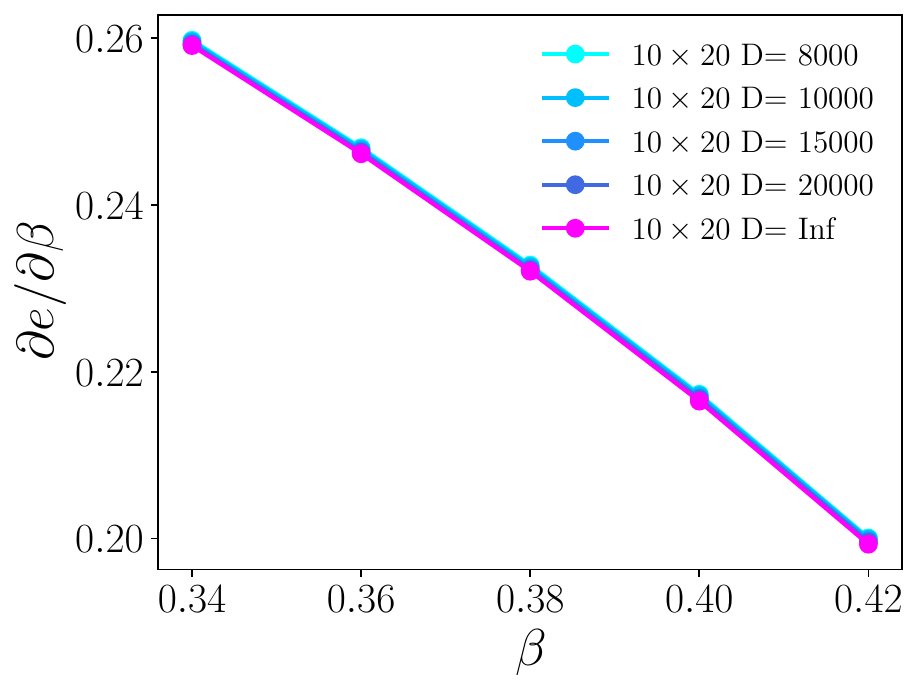}
    \includegraphics[width=70mm]{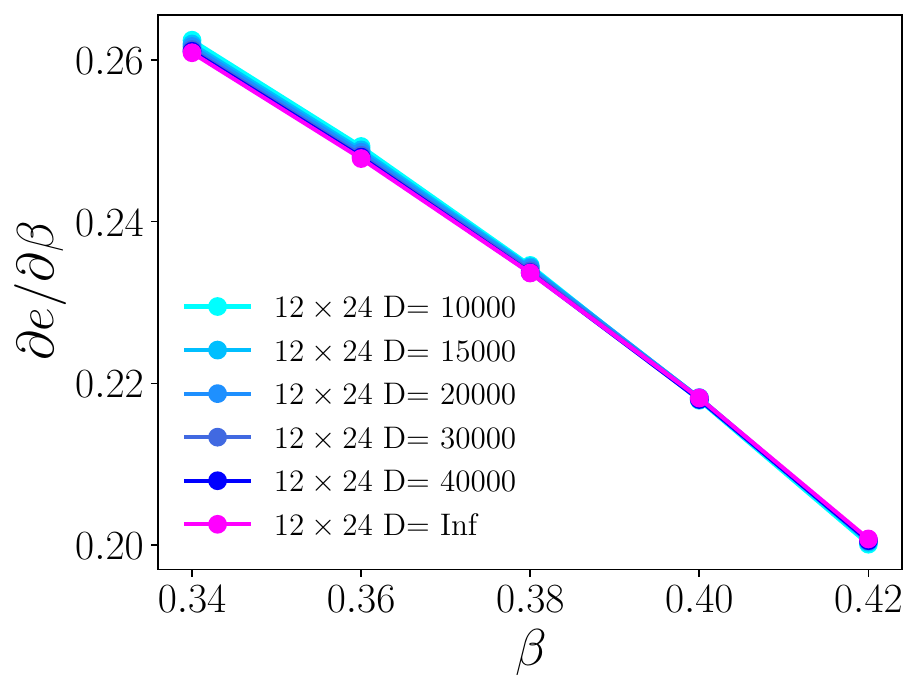}
    \includegraphics[width=70mm]{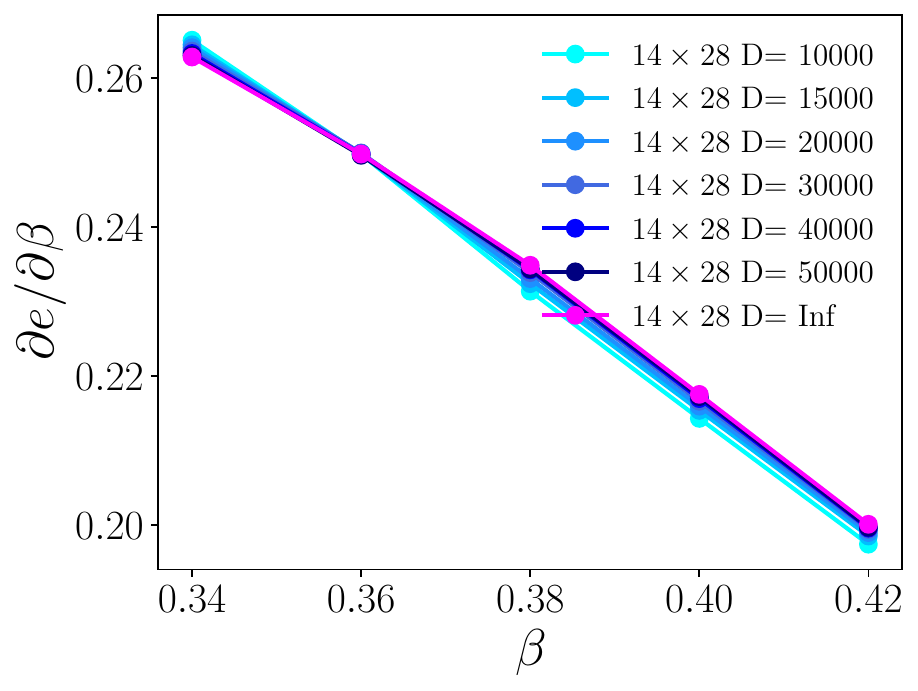}
    \caption{Convergence of the results for the first derivative of the ground state energy for $\Delta=0.4$}
    \label{dE_Delta4}
\end{figure*}

\begin{figure*}[t]
    \includegraphics[width=70mm]{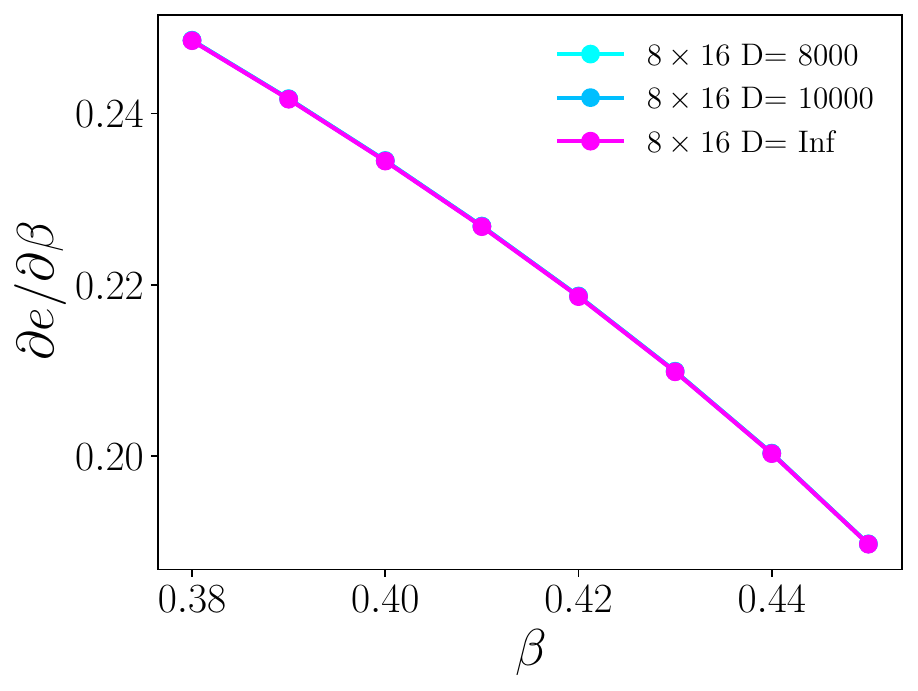}
    \includegraphics[width=70mm]{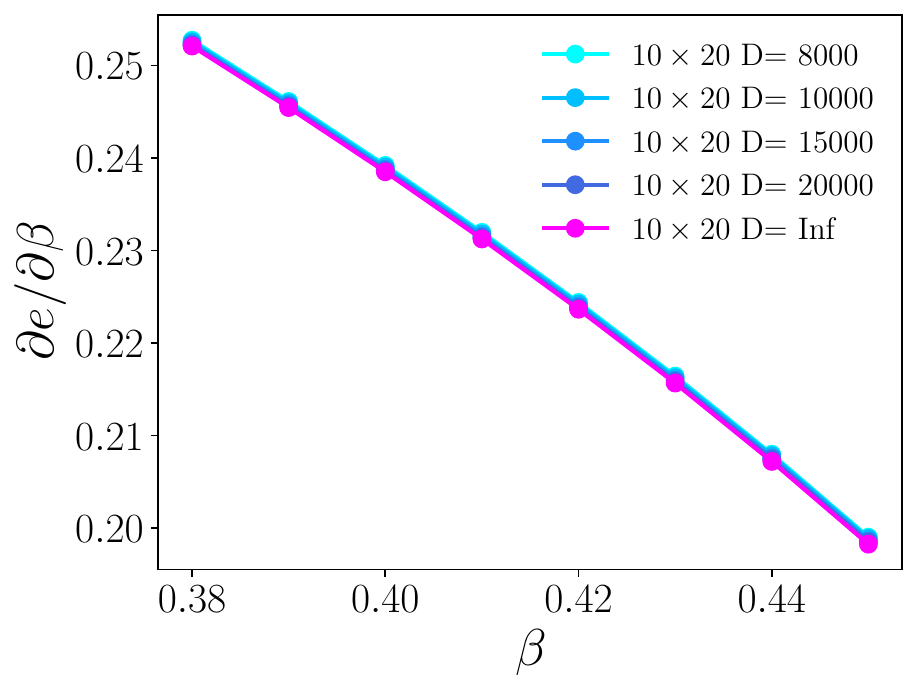}
    \includegraphics[width=70mm]{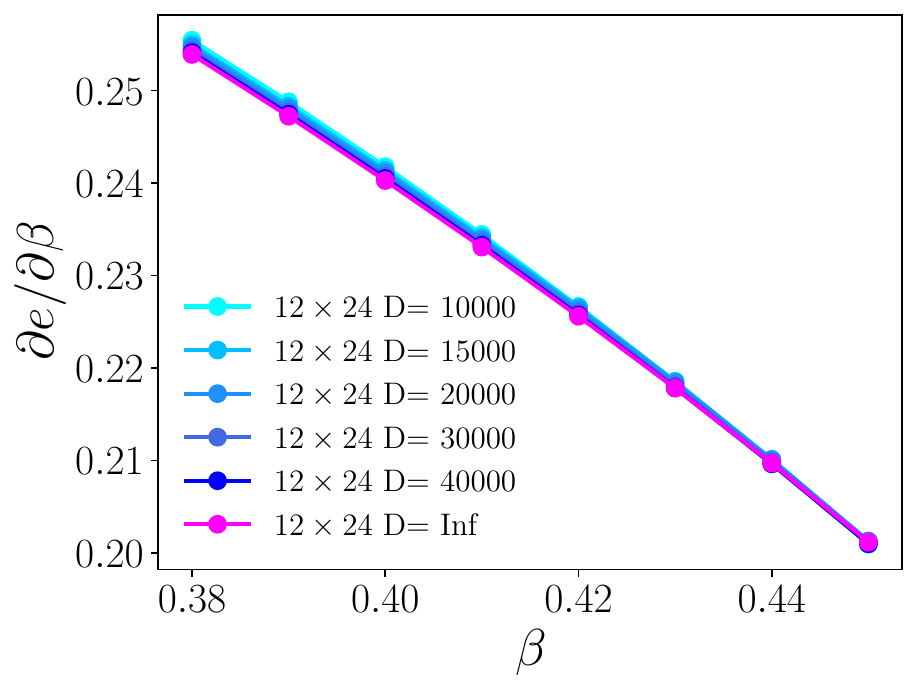}
    \includegraphics[width=70mm]{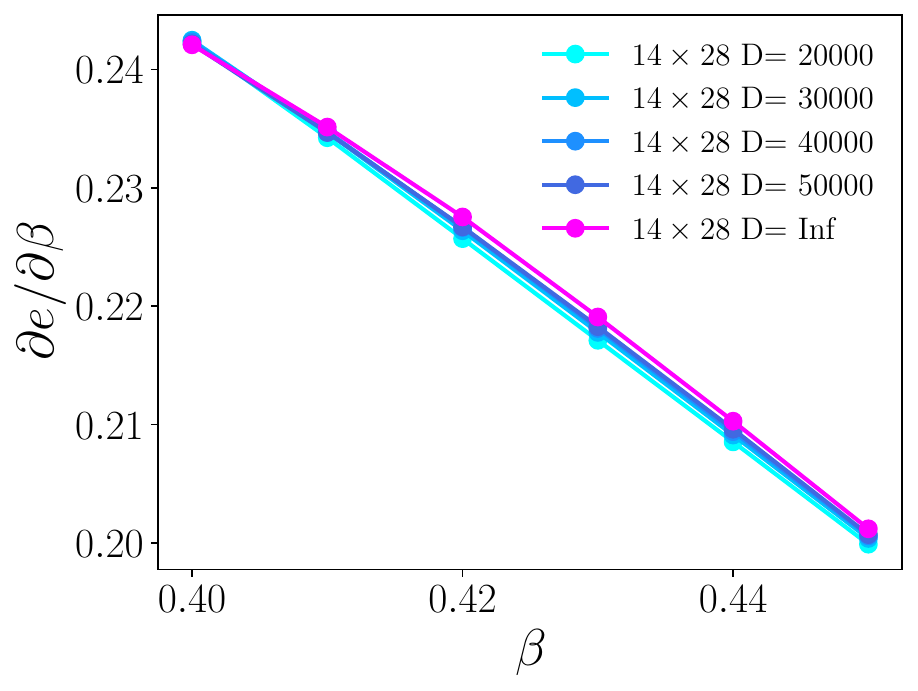}
    \caption{Convergence of the results for the first derivative of the ground state energy for $\Delta=0.3$}
    \label{dE_Delta3}
\end{figure*}

\begin{figure*}[t]
    \includegraphics[width=70mm]{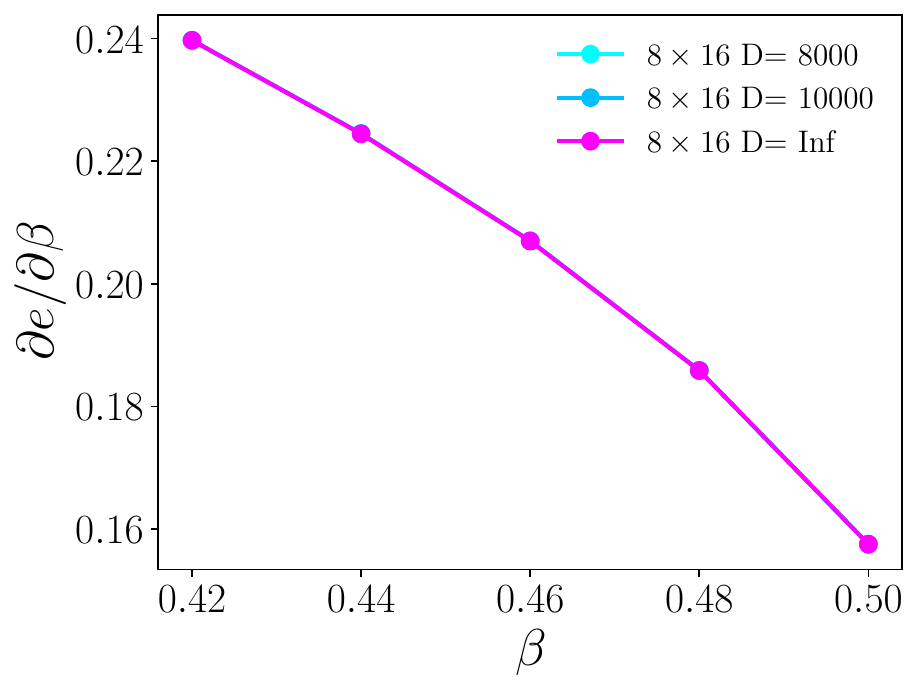}
    \includegraphics[width=70mm]{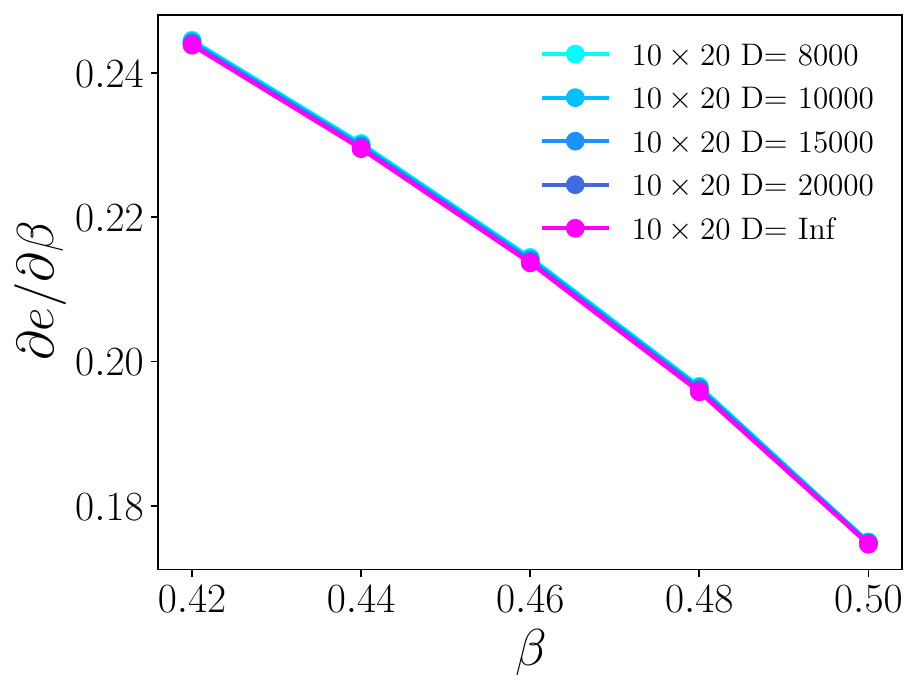}
    \includegraphics[width=70mm]{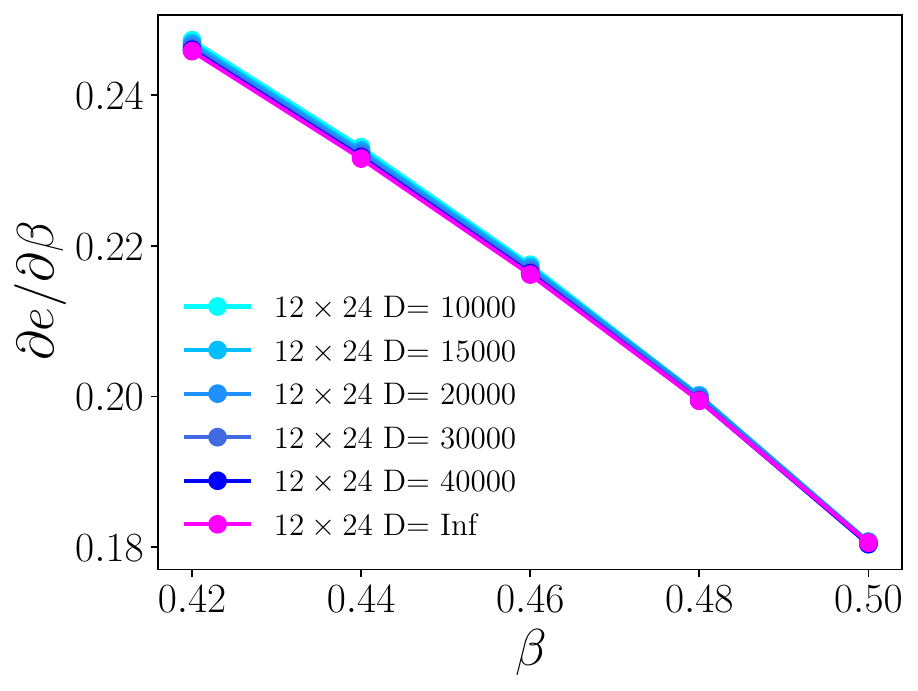}
    \includegraphics[width=70mm]{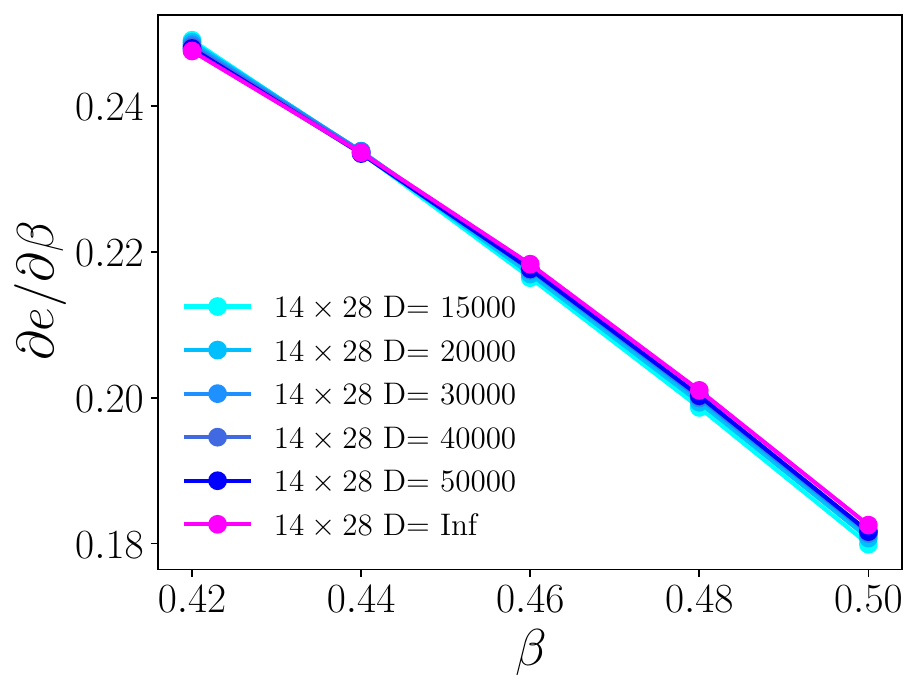}
    \caption{Convergence of the results for the first derivative of the ground state energy for $\Delta=0.2$}
    \label{dE_Delta2}
\end{figure*}

\begin{figure*}[t]
    \includegraphics[width=170mm]{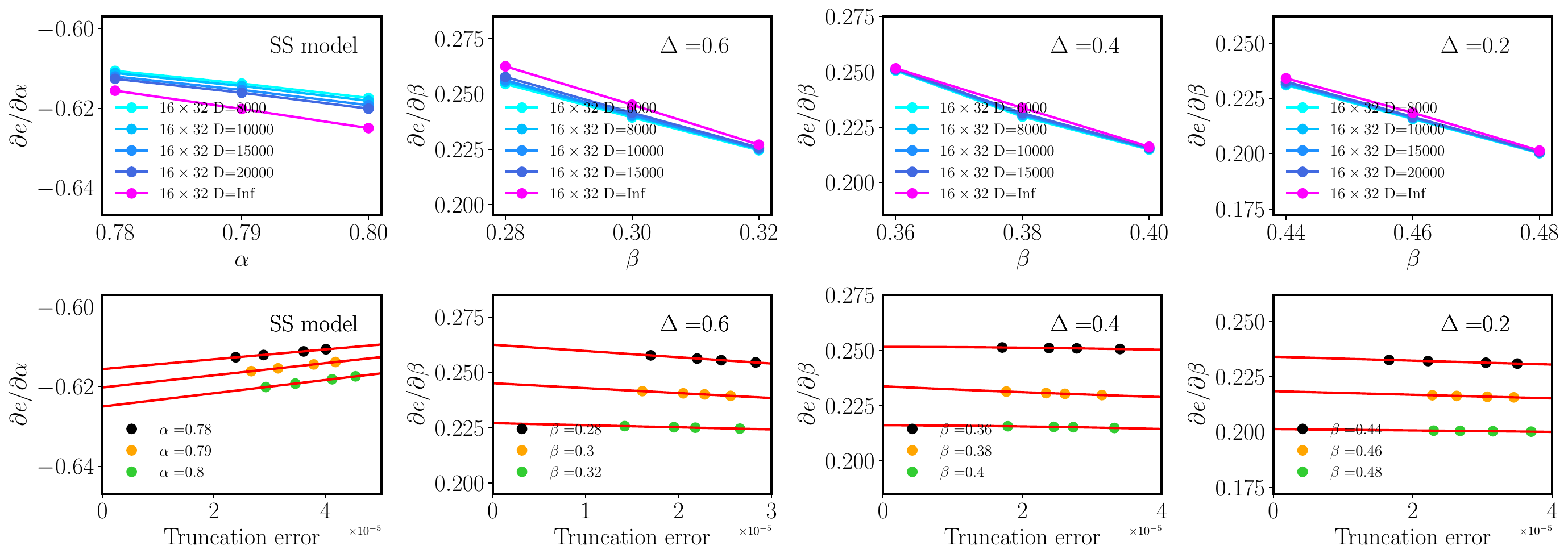}
    \caption{Convergence of the results for the first derivative of the ground state energy for the $16\times 32$ cylinders from linear extrapolation.}
    \label{dE_16}
\end{figure*}

For most of the calculations, the first derivative of the ground state energies converges rapidly with bond dimension. In Fig.~\ref{dE_SS}, \ref{dE_Delta6}, \ref{dE_Delta4}, \ref{dE_Delta3} and \ref{dE_Delta2}, where each figure is at different $\Delta$, the first derivative is linearly extrapolated as a function of truncation error. We also show the extrapolation results for $16\times 32$ systems in Fig.~\ref{dE_16}.

\subsection{The correlation ratio of the AFM order}

\begin{figure*}[t]
    \includegraphics[width=70mm]{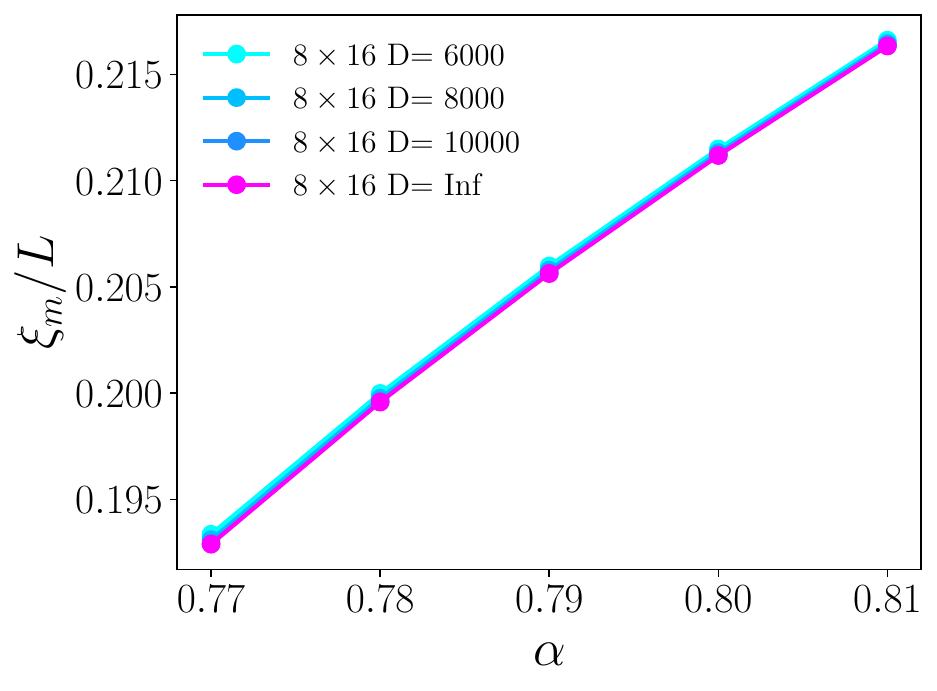}
    \includegraphics[width=70mm]{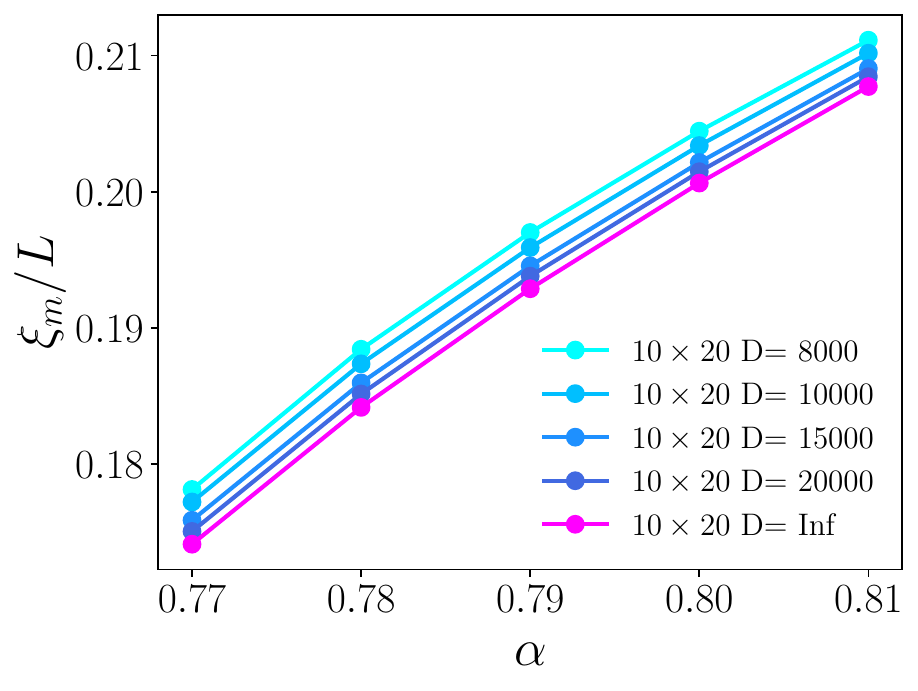}
    \includegraphics[width=70mm]{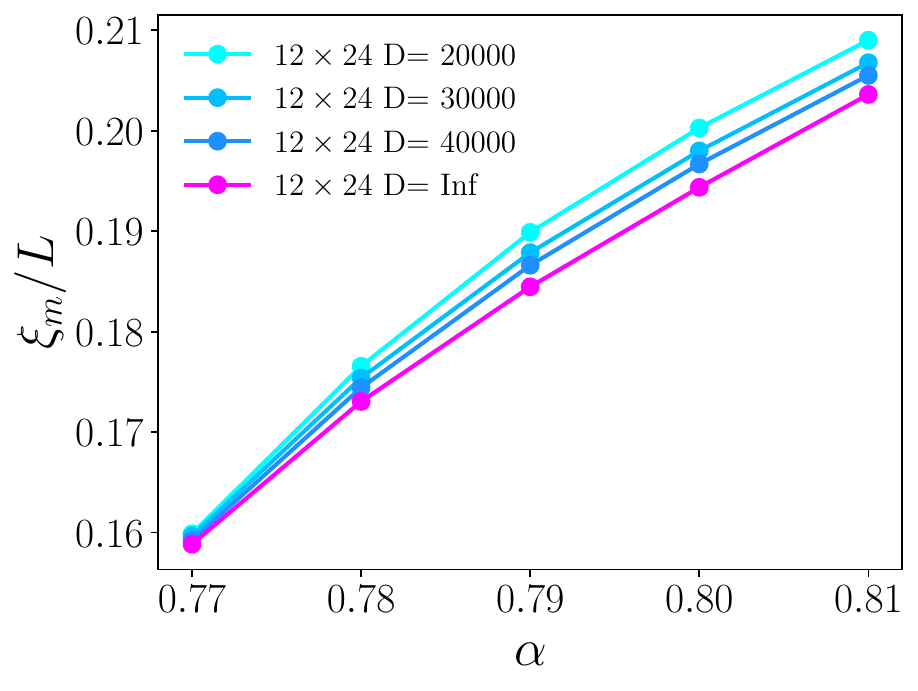}
    \includegraphics[width=70mm]{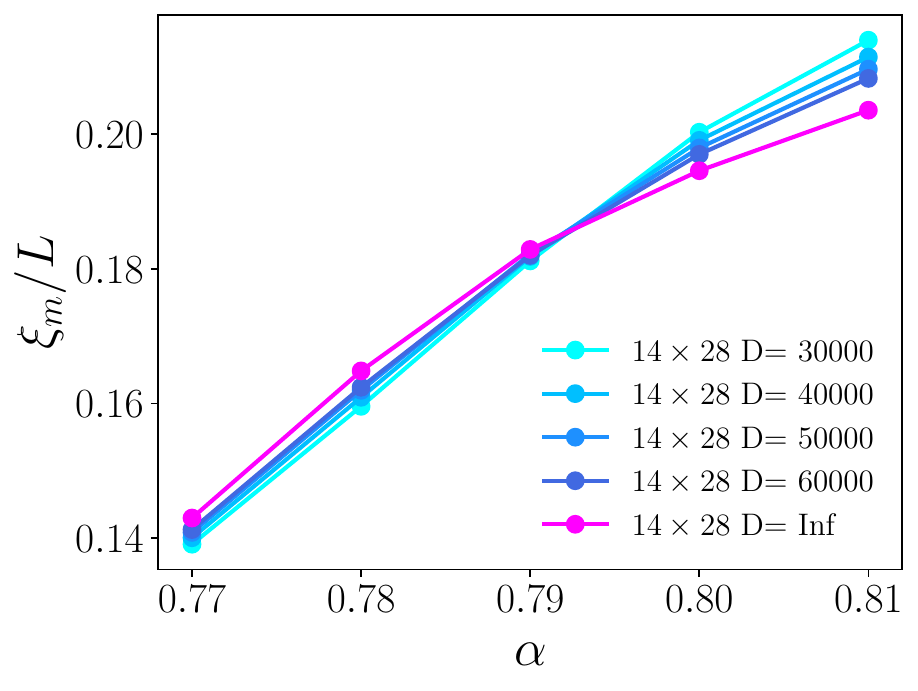}
    \caption{Convergence of the results for the correlation ratio for the Shastry-Sutherland model}
    \label{CORR_SS}
\end{figure*}

\begin{figure*}[t]
    \includegraphics[width=70mm]{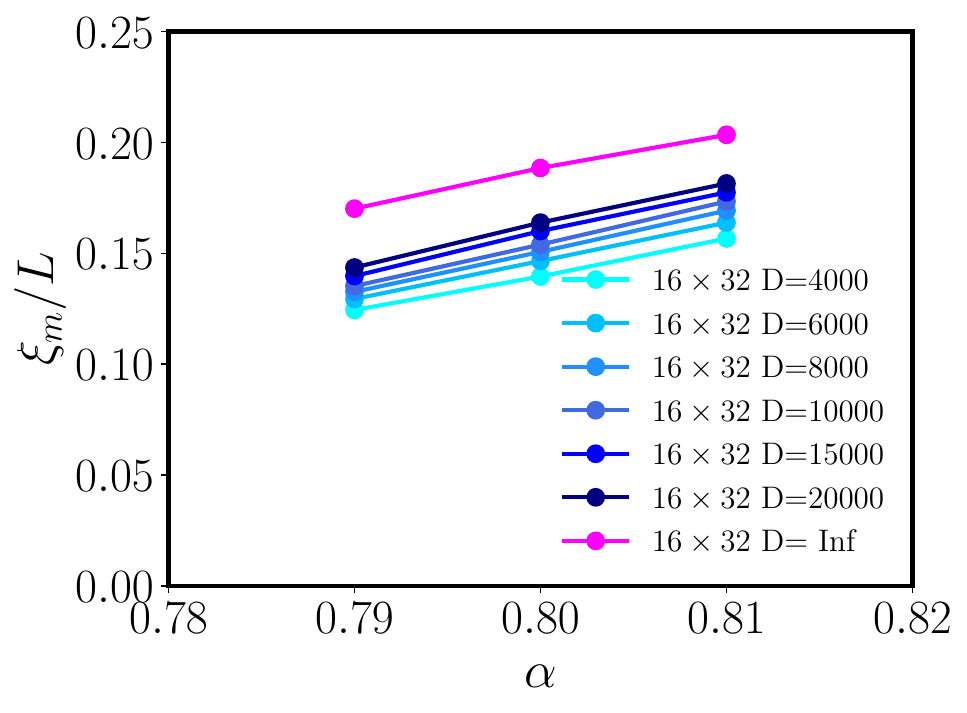}
    \caption{Convergence of the results for the correlation ratio for the SS model of $16\times32$ systems.}
    \label{CORR_16}
\end{figure*}

\begin{figure*}[t]
    \includegraphics[width=70mm]{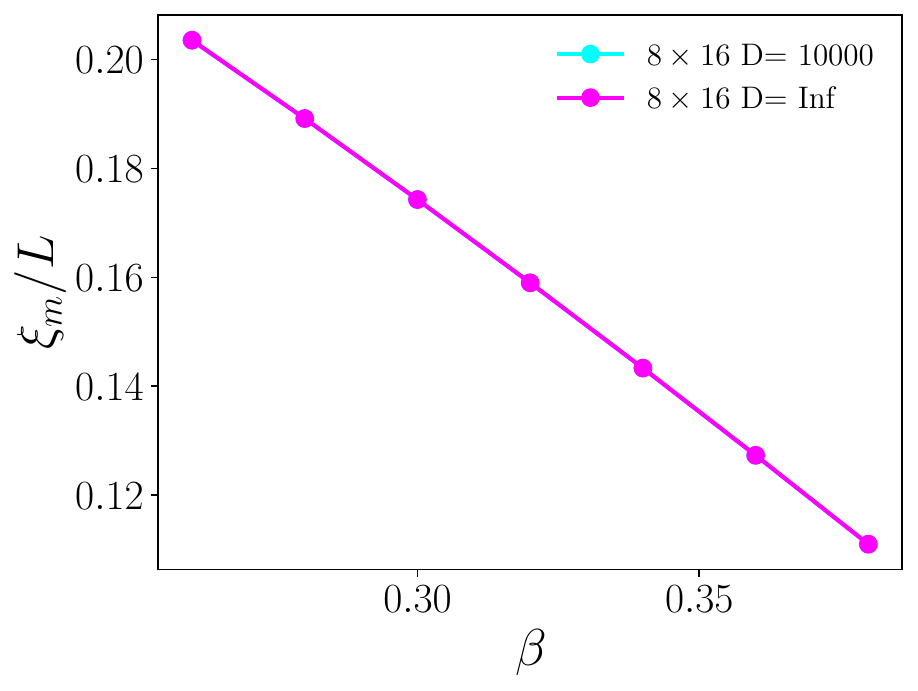}
    \includegraphics[width=70mm]{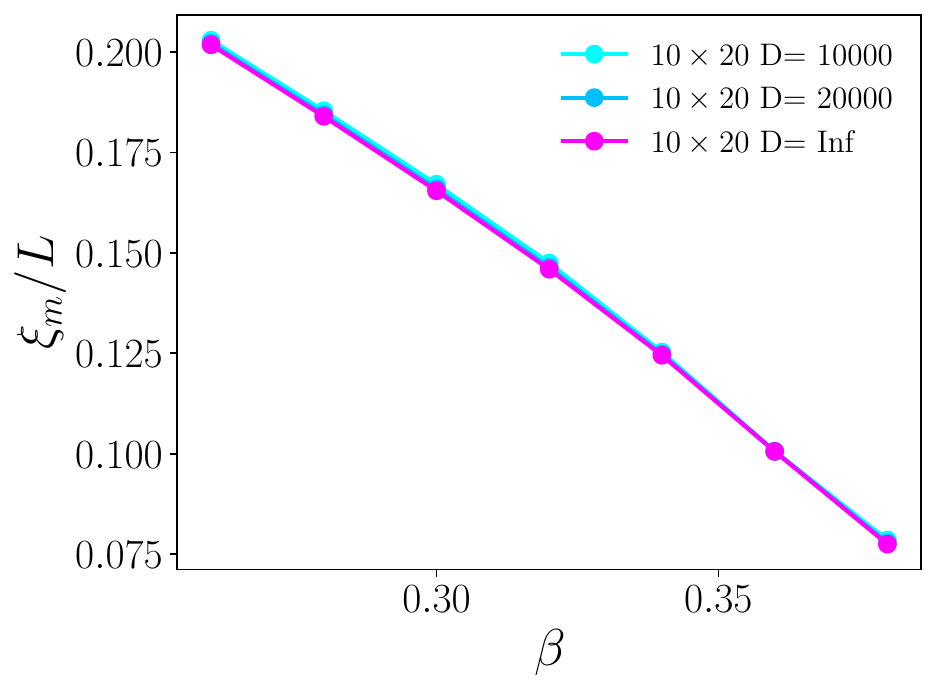}
    \includegraphics[width=70mm]{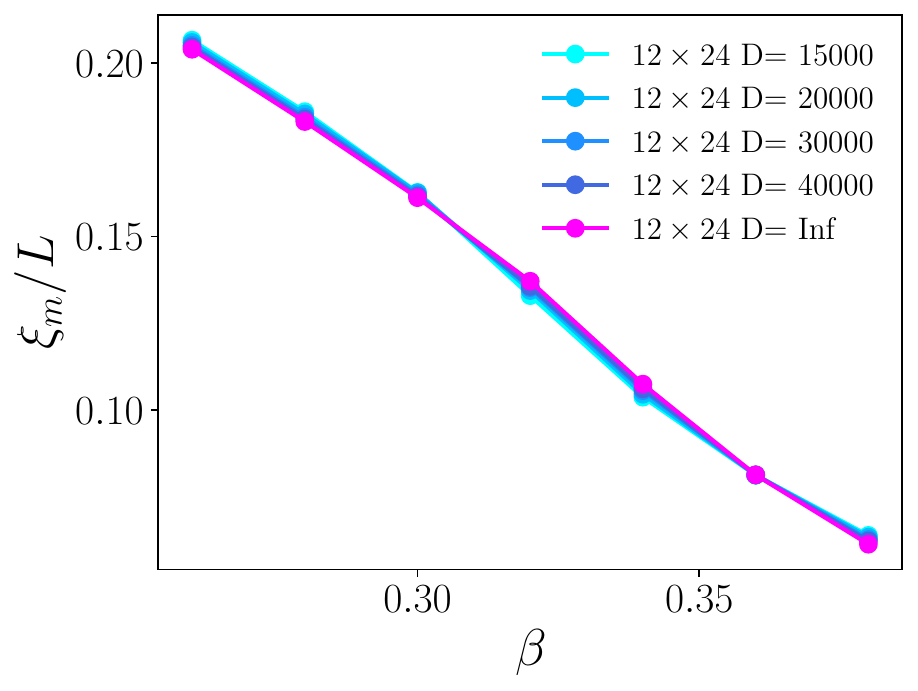}
    \includegraphics[width=70mm]{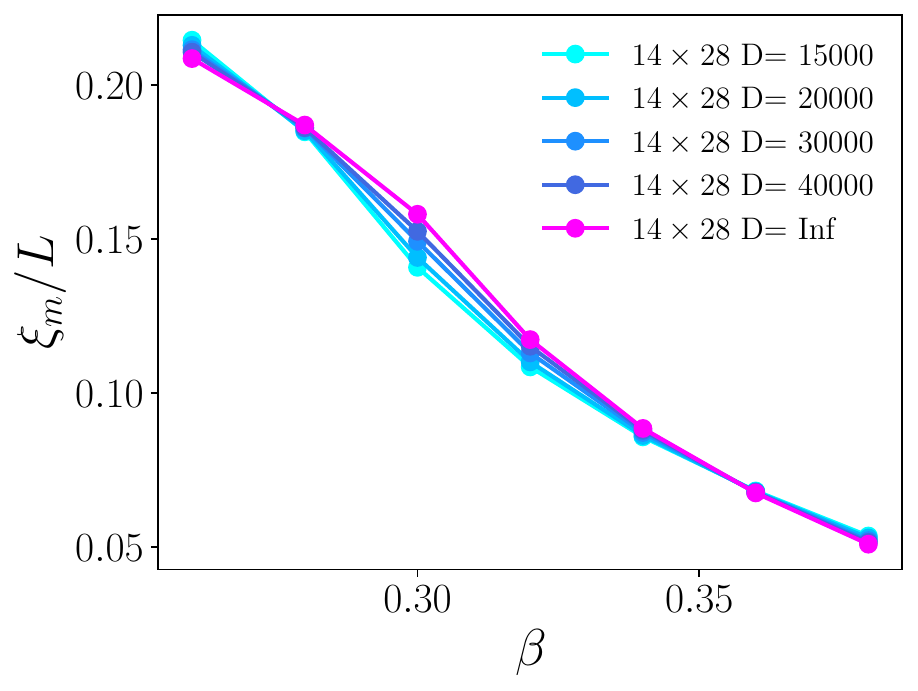}
    \caption{Convergence of the results for the correlation ratio for $\Delta=0.6$}
    \label{CORR_Delta6}
\end{figure*}

\begin{figure*}[t]
    \includegraphics[width=70mm]{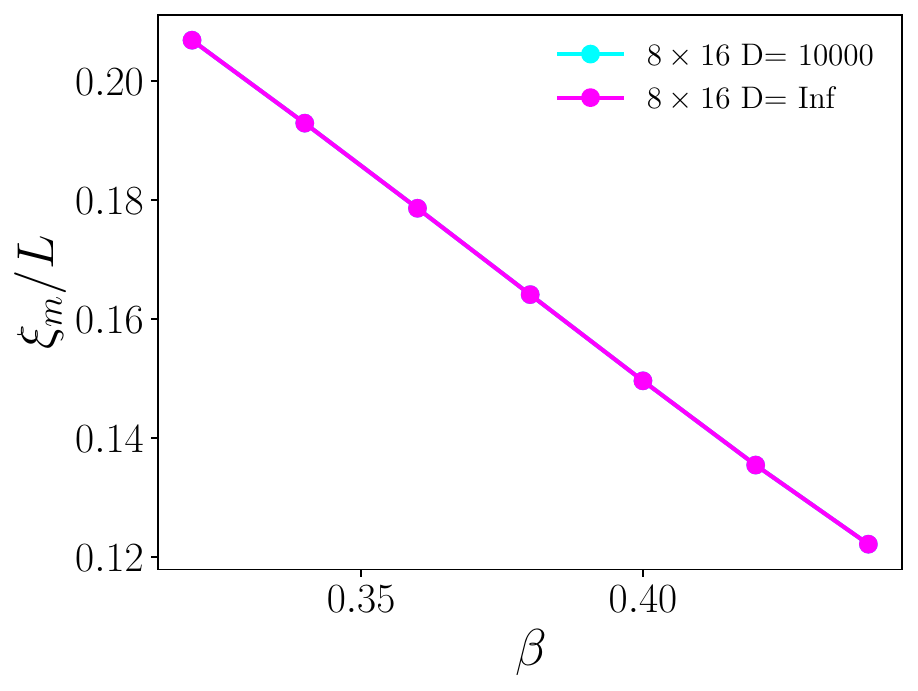}
    \includegraphics[width=70mm]{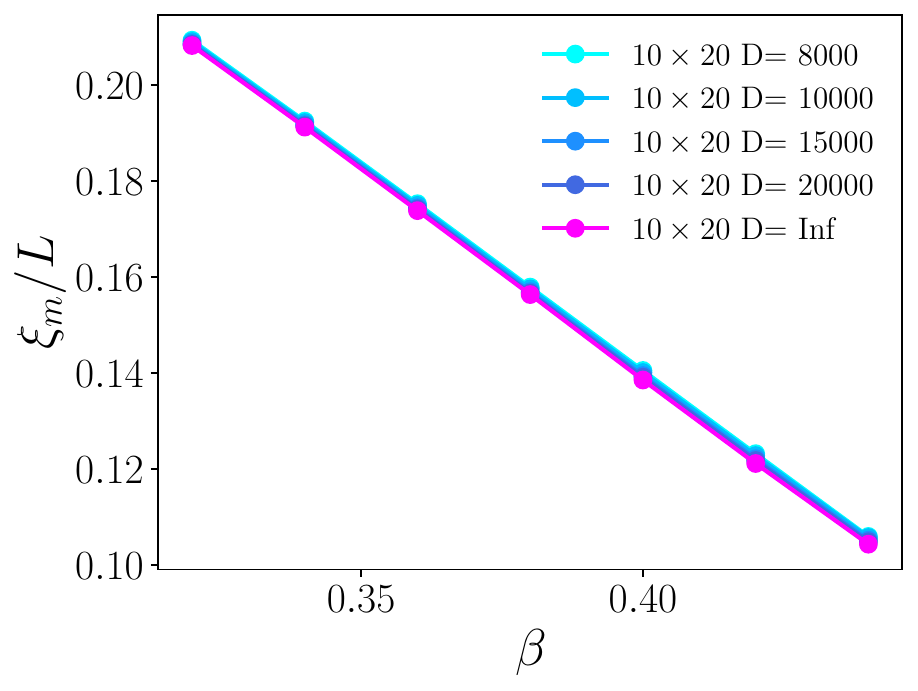}
    \includegraphics[width=70mm]{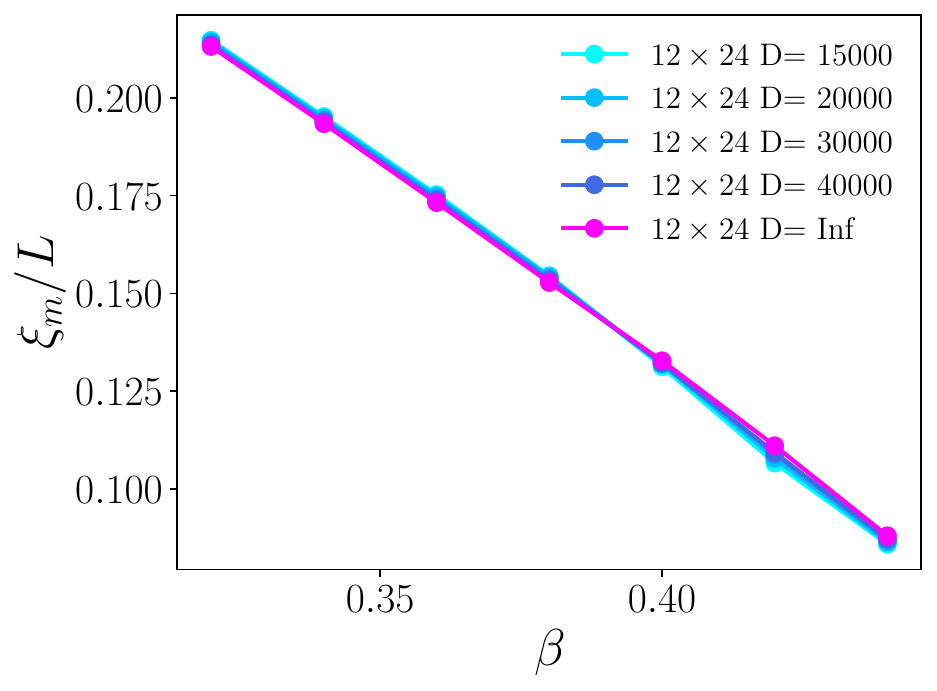}
    \includegraphics[width=70mm]{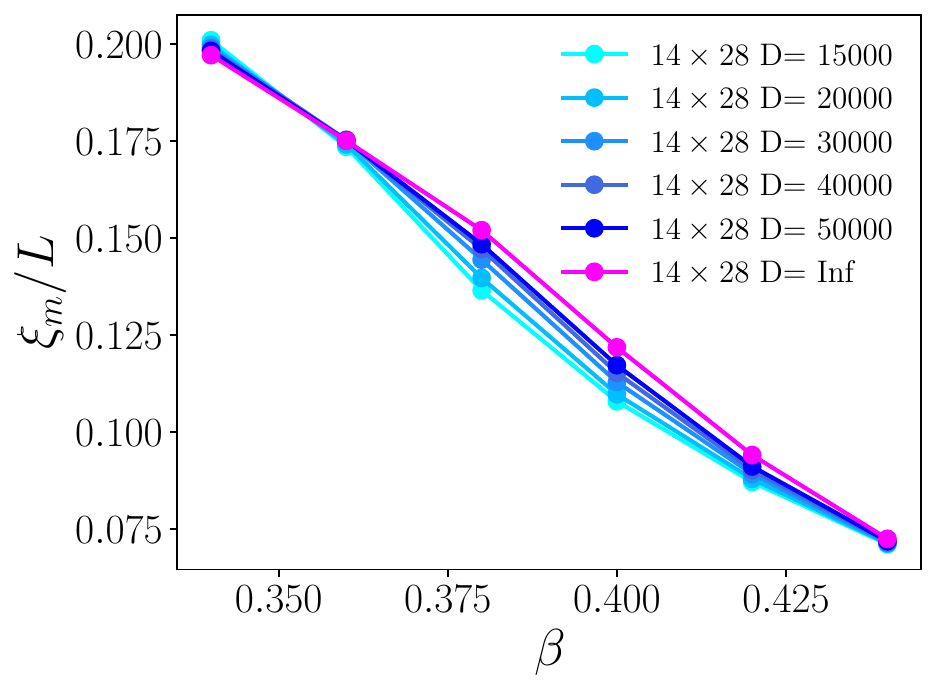}
    \caption{Convergence of the results for the correlation ratio for $\Delta=0.4$}
    \label{CORR_Delta4}
\end{figure*}

\begin{figure*}[t]
    \includegraphics[width=70mm]{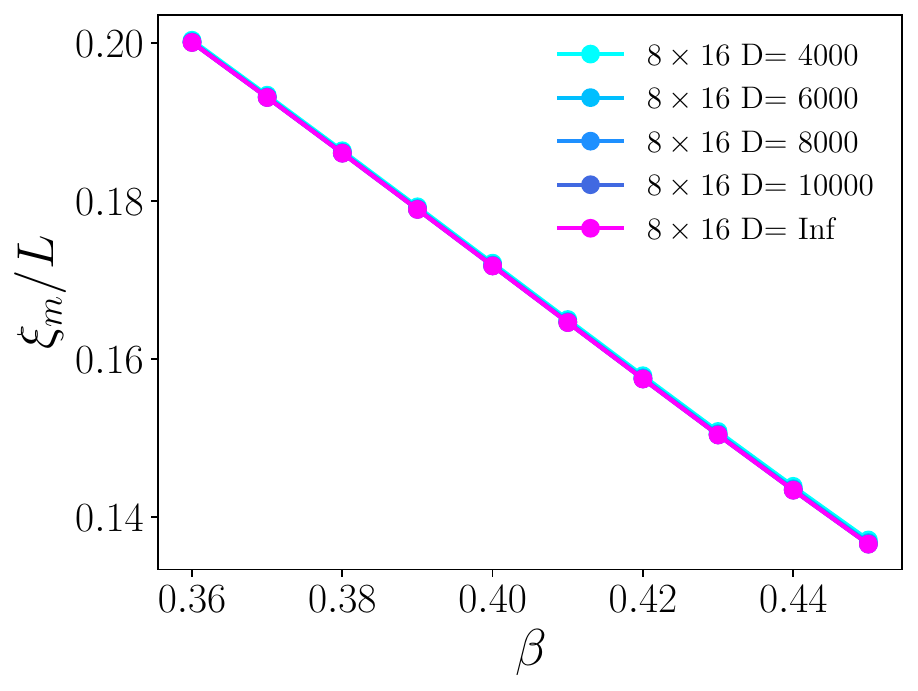}
    \includegraphics[width=70mm]{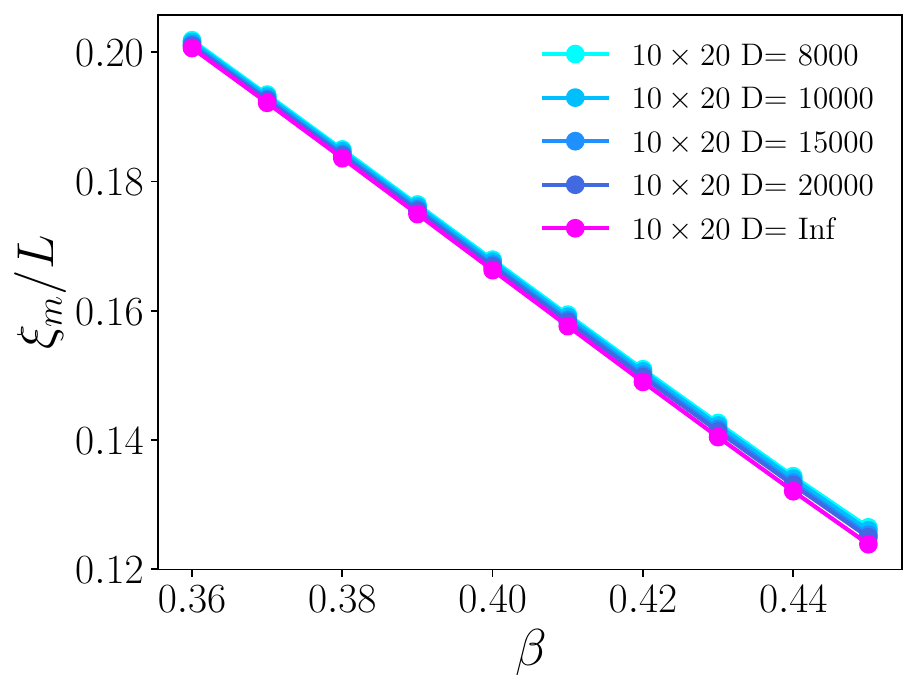}
    \includegraphics[width=70mm]{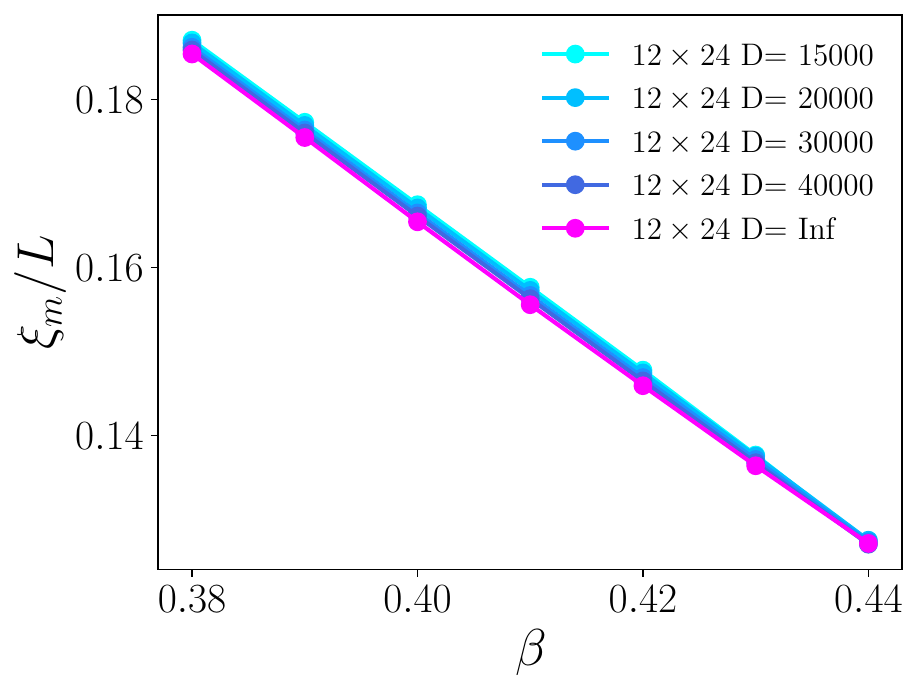}
    \includegraphics[width=70mm]{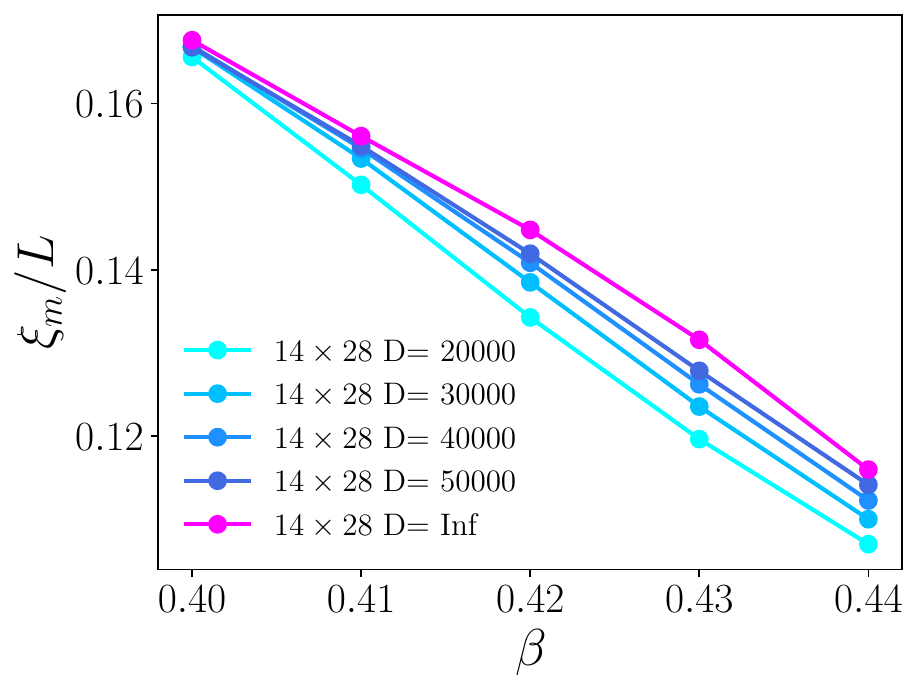}
    \caption{Convergence of the results for the correlation ratio for $\Delta=0.3$}
    \label{CORR_Delta3}
\end{figure*}

\begin{figure*}[t]
    \includegraphics[width=70mm]{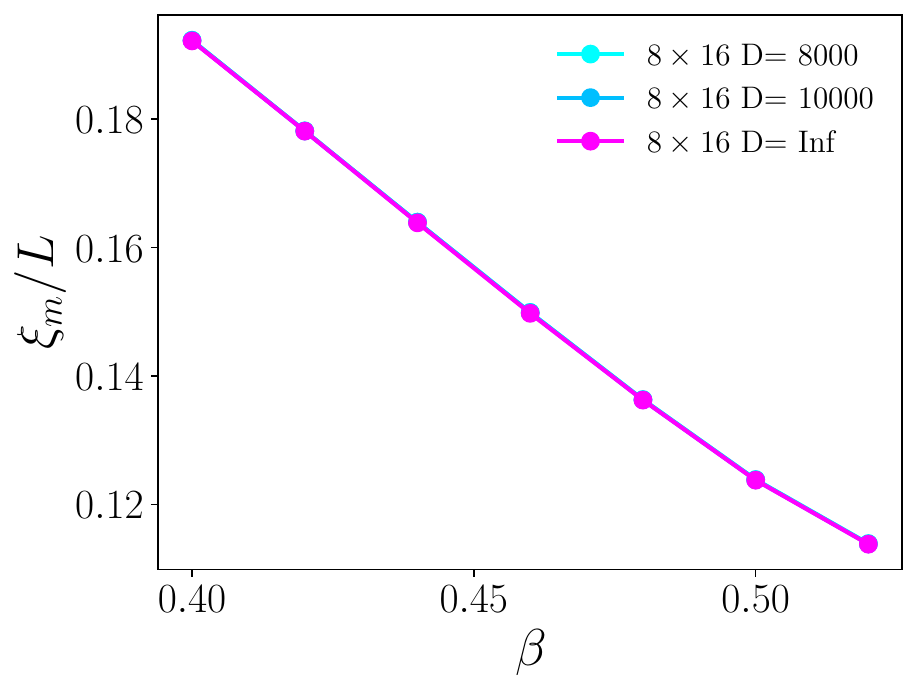}
    \includegraphics[width=70mm]{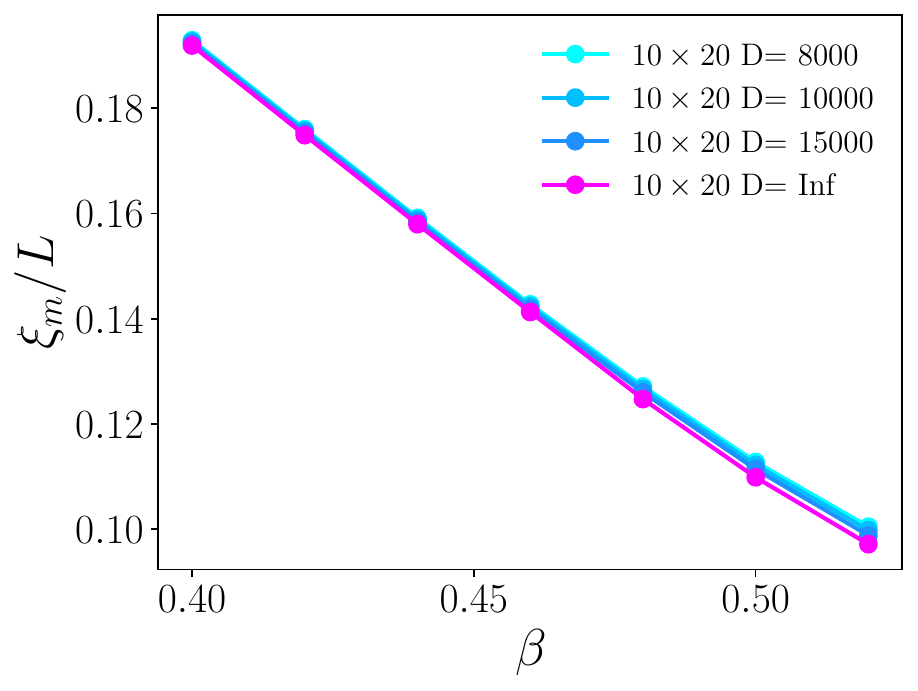}
    \includegraphics[width=70mm]{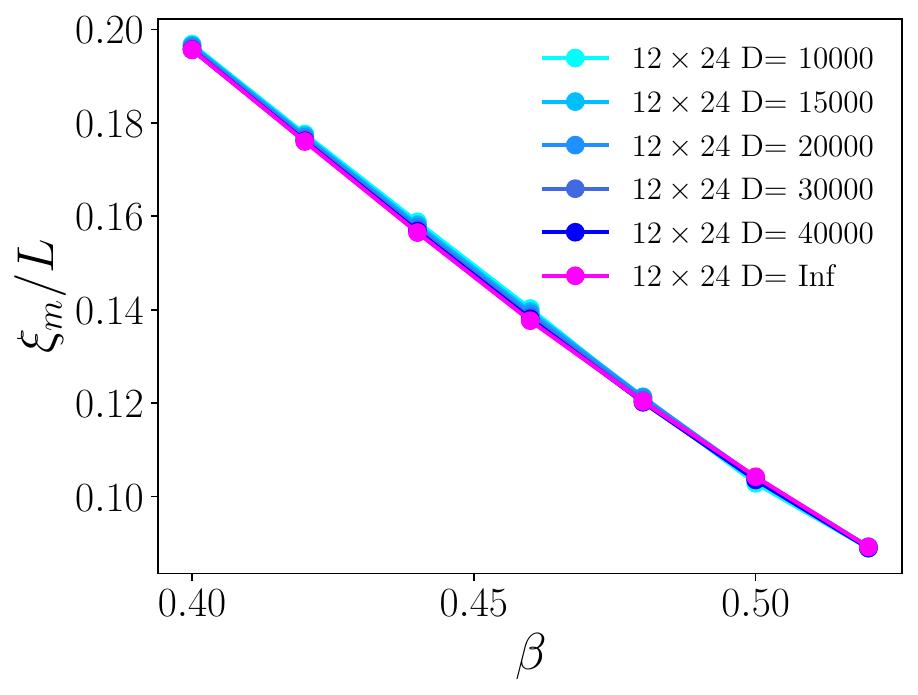}
    \includegraphics[width=70mm]{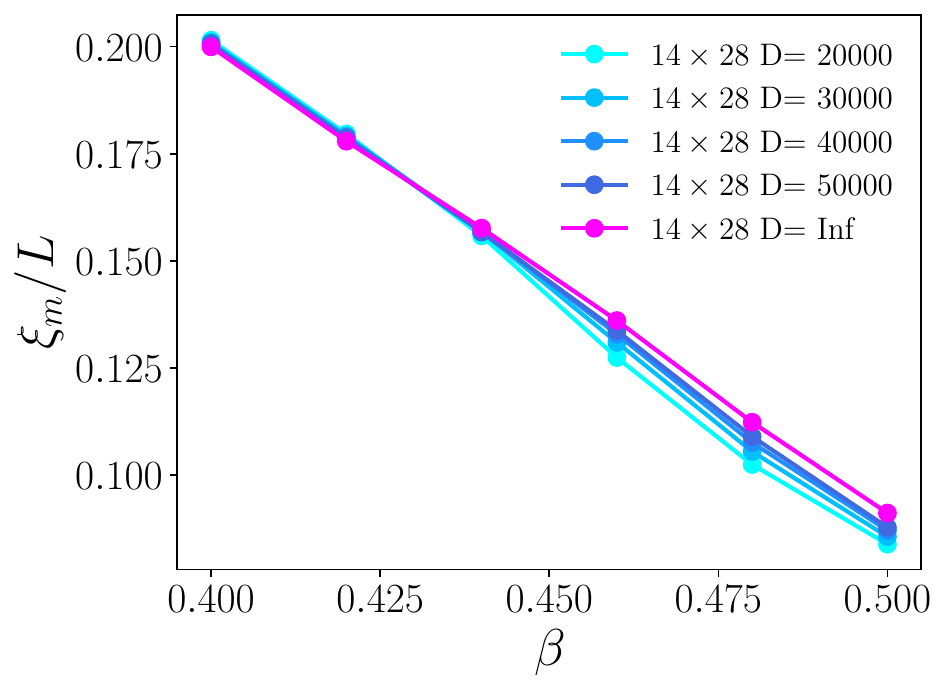}
    \caption{Convergence of the results for the correlation ratio for $\Delta=0.2$}
    \label{CORR_Delta2}
\end{figure*}

The correlation ratio for the AFM order parameter converges rapidly with the bond dimension, only exhibiting slower convergence near the transition point at very large system sizes. For most results, linear fits with truncation error are applied to reduce the finite bond dimension effect. For results with obvious curvature, second order polynomial fits with truncation error are applied. 
The results are shown in Fig.~\ref{CORR_SS}, \ref{CORR_16}, \ref{CORR_Delta6}, \ref{CORR_Delta4}, \ref{CORR_Delta3}, and \ref{CORR_Delta2}.


\end{document}